\documentclass[a4paper, 10pt, onecolumn]{article}      
\usepackage[top=1in, bottom=1.25in, left=1.25in, right=1.25in]{geometry} 
\usepackage[T1]{fontenc}
\usepackage[font=small,labelfont=bf,tableposition=top]{caption}

\usepackage{xfrac} 

\DeclareCaptionLabelFormat{andtable}{#1~#2  \&  \tablename~\thetable}

\usepackage{floatrow}
\newfloatcommand{capbtabbox}{table}[][\FBwidth]
\usepackage{blindtext}
\usepackage{epsfig} 
\usepackage{epstopdf}
\usepackage{gensymb} 
\usepackage{graphicx}     
\usepackage{amsmath}	

\usepackage{breqn} 

\usepackage{tabularx}

\usepackage{float}
\usepackage{subfig}
\usepackage{graphicx}

\usepackage{ulem} 
\allowdisplaybreaks		

\usepackage{url}

\usepackage{notoccite}

\title{\LARGE \bf
Controlled irradiation hardening of tungsten by cyclic recrystallization}

\author{A. Mannheim, J.A.W. van Dommelen and M.G.D. Geers}

\author{A. Mannheim, J.A.W. van Dommelen\footnote{Correspondence to J.A.W. van Dommelen. Electronic mail: j.a.w.v.dommelen@tue.nl}\hspace{0.1cm}  and M.G.D. Geers}

\date{Mechanics of Materials, Mechanical Engineering, Eindhoven University of Technology, P.O. Box 513, 5600 MB Eindhoven, The Netherlands \vspace{0.4cm}\\ \today}

\providecommand{\e}[1]{\ensuremath{\times 10^{#1}}}
\begin{document}

\maketitle
\thispagestyle{empty}

\begin{abstract}
The economical lifetime of the divertor is a key concern for realizing nuclear fusion reactors that may solve the world's energy problem. A main risk is thermo-mechanical failure of the plasma-facing tungsten monoblocks, as a consequence of irradiation hardening  induced by neutron displacement cascades. Lifetime extensions that could be carried out without prolonged maintenance periods are desired. In this work, the effects of potential treatments for extending the lifetime of an operational reactor are explored. The proposed treatments make use of  cyclic recrystallization processes that can occur in neutron-irradiated tungsten. Evolution of the microstructure under non-isothermal conditions is investigated, employing a multi-scale model that includes a physically-based mean-field recrystallization model and a cluster dynamics model for neutron irradiation effects. The model takes into account microstructural properties such as grain size and displacement-induced defect concentrations. The evolution of a hardness indicator under neutron irradiation was studied. The results reveal that, for the given microstructure and under the assumed model behaviour, periodical extra heating can have a significant positive influence on controlling the irradiation hardening. For example, at 800 \textdegree C, if extra annealing at 1200 \textdegree C was applied after every 100 hrs for the duration of 1 hr, then the hardness indicator reduces from maximum 140 to below 70.  
\end{abstract}

\section{Introduction}
In a tokamak-design nuclear fusion device, the component that receives the highest heat load (10 MW/m$^2$ for DEMO \cite{Bolt2004}) is the divertor. Monoblocks, made out of high purity tungsten, are positioned on the plasma-facing side of this component. They absorb the heat that was generated during the fusion reaction and direct it towards the cooling water. The monoblocks are actively cooled, and therefore the heat load induces a strong temperature gradient that leads to thermal stresses in the monoblock. Moreover, the tungsten is also damaged by high neutron and ion loads (leading to 15 dpa (displacements per atom) in 5 years of operation of DEMO and peak ion loads of 10$^{24}$ ions/m$^3$s \cite{Bolt2004}). Hereby, the ions mainly affect the surface of the monoblock, while the neutrons affect the entire monoblock. Despite these extreme conditions, the divertor needs to have a minimum lifetime of 2 years prior to replacement for reasons of economical viability. This means that interim repairs for prolongation of the divertor lifetime are most likely needed.\\
\\
The lifetime of the tungsten monoblocks under heat and neutron irradiation will depend strongly on the evolution of the strength and ductility of the material. A high hardness, which is reported for fission-neutron-irradiated tungsten \cite{Hu2016,Fukuda2016} can be associated to a high yield strength and low ductility. \\
\\
The high density of displaced atoms (resulting from the continuous neutron displacement cascades that are generated by the 14.0 MeV fusion neutrons) entails a high stored energy in the material. The ongoing accumulation of stored energy, in combination with sufficiently high temperatures, can lead to (repeated) dynamic recrystallization. During dynamic recrystallization, the material's microstructure is completely renewed, from heavily defected to virtually free of lattice defects. Dynamic recrystallization can be instrumental in decreasing the material's hardness. In \cite{Dini2010}, it was shown for steel that (strain-induced) recrystallization can lead to softening and to an improved ductility. A treatment that achieves an improved ductility and removes the accumulated lattice damage, is most likely necessary for neutron-irradiated tungsten. Although, depending on the composition, recrystallization can lead to embrittlement, under the right circumstances, it may lead to an improved ductility by regeneration of the lattice.\\
\\
For most materials, recrystallization leads to an increase in ductility and a decrease in yield strength. For tungsten, the literature is divided on this point. Degradation of the ductility for post-recrystallization was reported in \cite{Reiser2012} and \cite{Terentyev2015}. However, the testing temperature may have been too low in that analysis to improve the ductility (below the brittle-to-ductile transition temperature, which is reported to increase by recrystallization).  On the other hand, there are also examples in literature where the ductility of tungsten increases after recrystallization, such as by Wirtz \textit{et al.} \cite{Wirtz2017}, who found for several grades of tungsten a significantly improved uniform elongation and total elongation in post-recrystallization (accompanied by a lower yield strength), or \cite{Wirtz2017PS}, where  fracture strains were reported of 17\% and 22\% prior to recrystallization and 68\% after recrystallization. The impurity content of the tungsten grade may play a role in the grain boundary cohesion \cite{Scheiber2016}, which can affect the ductility. Recrystallization may lead to an increase in impurity concentrations at the grain boundary surfaces, because the impurities are redistributed and because the grain boundary surface density changes for a different average grain size. It was suggested that the ductility in tungsten may be determined by the presence of edge dislocations \cite{Ren2018}, but the mechanisms that control the ductility of tungsten are not properly understood, whereby the effects of grain size and cold working were not studied separately yet. \\
\\
The combined effect of heat and neutron load on the microstructural evolution of tungsten were already studied in \cite{Mannheim2018, Mannheim2018b}, using multi-scale modelling. In \cite{Mannheim2018}, a multi-scale model was presented, consisting of a mean-field recrystallization model and a cluster dynamics model. The model was parametrized based on static recrystallization experiments \cite{Lopez2015} and the evolution of the microstructural properties (grain size distribution, defect distribution) was predicted. Cyclic neutron-induced recrystallization was predicted to occur for temperatures in the range 1000\textdegree C - 1400\textdegree C \cite{Mannheim2018b} for the considered microstructure, displacement damage rate and material behaviour. In addition, only a limited irradiation hardening was predicted for cyclic recrystallization \cite{Mannheim2018b} at high temperatures.  Neutron-induced recrystallization and grain growth were also studied in \cite{Vaidya1983, Kaoumi2008}.\\
\\
In this paper, the main objective is to explore the use of dedicated interim heat treatments of tungsten for manipulating the microstructure such that the divertor lifetime may be prolonged.  In particular, it is studied whether temporary, periodical, extra heating can be used to keep the hardness increase below a certain level. For these non-isothermal studies, the multi-scale model that was presented in \cite{Mannheim2018b} is extended, employing a multitude of homogeneous equivalent media (HEMs) instead of just two HEMs, to eliminate the effects of additional parameters in the grain growth part of the model. Furthermore, because of these extensions, a reparametrization of the model is done and it is discussed which recrystallization experiments would be valuable in order to further improve model predictions. 

\section{Method}

\subsection{Multi-scale approach}
\begin{figure}[H]
\includegraphics[width=0.7\textwidth]{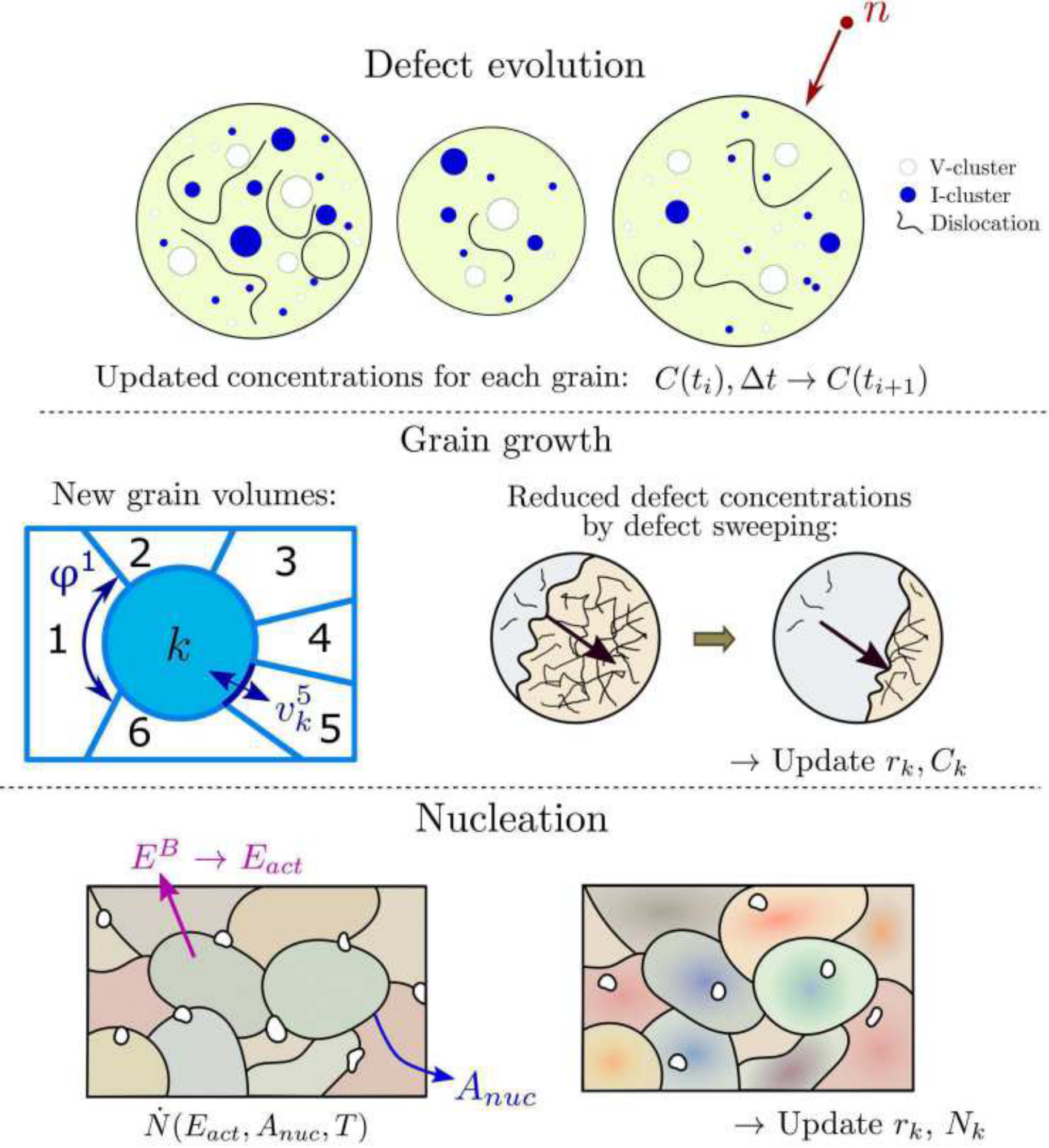}
\caption{Schematic overview of the multi-scale model, showing the key mechanisms (defect evolution, grain growth and nucleation) and their impact on the microstructural evolution. Here, $C$ is the defect concentration, $\Delta t = t_{i+1} - t_i$ is the size of the $i^{th}$ time increment, $\phi^1$ is the grain boundary surface fraction that is shared between HEM 1 and each representative grain $k$, $v_k^5$ is the velocity of the grain boundary segment that is shared between HEM 5 and grain $k$, $r_k$ is the radius of grain $k$, $C_k$ is the concentration of a defect in grain $k$, $E^B$ is the bulk stored energy density, $E_{act}$ is the activation energy for nucleation, $A_{nuc}$ is the nucleation surface area, $\dot{N}$ is the nucleation rate, $T$ is the temperature and $N_k$ is the number of grains that is represented by grain $k$.}
\label{fig:scheme}
\end{figure}

A multi-scale model is used to describe microstructural evolution under neutron loading for a prescribed temperature profile (isothermal or non-isothermal). The framework of the model was described elaborately in \cite{Mannheim2018b} and is further extended here. A schematic representation of the model is shown in Figure~\ref{fig:scheme}. The evolution of a set of (spherical) grains that is representative for the microstructure (hereafter referred to as representative grains), is predicted. For each grain, the evolution of the defect concentrations (in the form of dislocations, vacancies, self-interstitial atoms and their clusters) is predicted using Cluster Dynamics \cite{HardouinDuparc2002, Li2012}. The lattice damage that accumulates induces a bulk stored energy $E^B$ in the microstructure, which is strongly temperature dependent and which may drive recrystallization. The nucleation of new, nearly defect-free grains and grain growth are described using a mean-field recrystallization model based on \cite{Bernard2011}. In this model, each representative grain interacts with surrounding averaged media (homogeneous equivalent media, HEMs). The amount of growth of grain $k$ is determined by the differences in stored energy $\Delta E^{HEM}_k$ between each of the HEMs and the grain. As the grain boundaries move, the encountered defects are swept, so that grain growth affects both the defect concentrations and the grain radii. Two types of nucleation can occur in the model: bulk and necklace nucleation. The amount of nucleation that occurs at a certain time, depends on the nucleation activation barrier $E_{act}$ (which depends on the amount of stored energy in the microstructure) and on the nucleation surface area $A_{nuc}$ or nucleation volume $V_{nuc}$, available for necklace and bulk interaction, respectively. The grain boundary mobility $m$, which is strongly temperature dependent, plays an important role in the rates of grain growth and nucleation. The model is solved incrementally, in a staggered way, according to Figure~\ref{fig:scheme}.

\subsection{Cluster dynamics}
Neutron-induced displacement cascades result in the formation of many vacancies ($V$), self-interstitial atoms ($I$) and clusters of these ($V_n$, $I_n$, where $n$ refers to the amount of vacancies or self-interstitial atoms contained in a cluster). Indirectly, the displacement cascades affect the dislocation density ($\rho$) as well: by climb and by Bardeen-Herring sources  \cite{Stoller1990}. The evolution of the concentrations of the defect clusters is described by a set of coupled rate equations \cite{Li2012,HardouinDuparc2002}, that consist of production ($G$), reaction ($J$) and annihilation ($L$) terms: 
\begin{alignat}{2}
&\frac{d}{dt} \bigg[ C_{I_n} \bigg]  	&&  	=  G_{I_n} +  J^I_{n-1,n} + J^I_{n+1,n} - [J^I_{n,n-1} + J^I_{n,n+1}]  - L_{I_n} C_{I_n} , \label{eq:gen1}\\
&\frac{d}{dt}\bigg[ C_{V_n} \bigg]  	&&  	= G_{V_n} +  J^V_{n-1,n} + J^V_{n+1,n} - [J^V_{n,n-1} + J^V_{n,n+1}]  - L_{V_n} C_{V_n} , \label{eq:gen2}\\
&\frac{d}{dt}\bigg[ \rho \bigg]    	&&      = 2 \pi v_{cl} S_{BH}  - \rho \tau_{cl}^{-1} . \label{eq:gen3}
\end{alignat}Here, $C_{I_n}$, $C_{V_n}$ and $\rho$ are the concentrations of self-interstitial clusters $I_n$, vacancy clusters $V_n$ and the density of dislocations, respectively, and $L$ represents the dislocation and grain boundary sinks for the mobile defects. $J^I_{n,n+1}$ represents the rate of the reactions $I_n + I \rightarrow I_{n+1}$ and $I_n \rightarrow I_{n+1} + V$ and analogous for the other $J$. Large $I_n$-clusters can become part of the dislocation network and the fraction of the growing clusters of size $n$ that does that is denoted by $f_n$ \cite{Jourdan2015}. Dislocations can be generated by Bardeen-Herring sources and they can be removed by dipole annihilation. $v_{cl}$ denotes the climb velocity, $S_{BH}$ the density of Bardeen-Herring sources, and $\tau_{cl}$ the average dislocation lifetime prior to dipole annihilation. In the model, $I_1$ and $V_1$ are considered mobile. These are the defects that make absorption, emission and annihilation at sinks possible. The detailed expressions are given in Table~\ref{tab:parval} in Appendix~\ref{app:cd}. \\
\\
Note that the cluster dynamics model has a temperature-dependence and a grain size dependence. The temperature dependence can affect the absorption and emission rates (which depend on the diffusion of the mobile defects) and the climb velocity of the dislocations. The grain boundary sink strength depends on the grain size: in larger grains, the mobile defects have more difficulty to reach the grain boundaries. However, depending on the temperature, the smaller grains may be the fastest to accumulate damage, as the evolution of the defect concentrations are all interconnected.

\paragraph{Damage production}\mbox{}\\
The temperature-dependent power law for the defect generation rates during displacement cascades was discussed before in \cite{Mannheim2018,Mannheim2018b}. In this work, the expression is further refined to:
\begin{align}
G_{\epsilon_n} = \frac{(1-f_D/f_{max})\eta A_{\epsilon}}{n^{S_{\epsilon}}}.
\end{align}
Here, $\epsilon=I$ or $V$ denotes the defect type, $G_{\epsilon_n}$ is the damage production rate of defects of size $n$ and type $\epsilon$, $S_{\epsilon}$ and $A_{\epsilon}$ are temperature-dependent parameters, with $A_{\epsilon}=G_0 / \sum_{n=1}^{N_{max}} n^{1-S_{\epsilon}}$ For intermediate temperatures, the parameter values are interpolated using the values of Table~\ref{tab:ownpowerlaw}, which are based on MD-simulation results. Further, $\eta$ is a parameter that denotes the fraction of defects, out of those that have survived the MD-simulations, that remain on longer time scales \cite{Mannheim2018b}. In this work, $\eta=1$ is used, but similar simulations results (at different time scales) may be expected if a value of $\eta=0.01$ would be used, based on limited simulations. Finally, the term $(1-f_D/f_{max})$ ensures a decrease in defect production rate as damage accumulates in the material, where $f_D$ denotes the atomic defect fraction, and $f_{max}$ denotes the saturated defect fraction level.

\begin{table}[ht!]
\centering
    \begin{tabular}{| l | l | l | l | }
    \hline
	$T$ (K)				&	$G_0$ ($\#$ defects/atom s)	&	$S_I$ 		&	$S_V$	 	\\ \hline
	300					&	4.3\e{-8}				&	2.20		&	1.63 		\\ \hline
	1025					&	3.3\e{-8} 				&	2.50		&	1.86 		\\ \hline
	2050					&	3.1\e{-8} 				&	2.17		&	2.42 		\\ 
        \hline
    \end{tabular}
    \caption{Parameter values for the defect production rate $G_0$ and power law exponents $S_I$ and $S_V$ at different temperatures, from \cite{Mannheim2018}, based on MD-results \cite{Setyawan2015}. }
    \label{tab:ownpowerlaw}
\end{table} 

Few studies are available regarding the defect saturation level in bcc metals. In 1996, Gao and Bacon studied the effect of overlapping cascades on the microstructural evolution for bcc-Fe, using MD-simulations \cite{Gao1995}. Recently, Sand \textit{et al.} \cite{Sand2018} showed that the microstructural evolution due to overlapping cascades strongly differs from metal to metal. For the bcc metals Fe and W, subcascades easily form in Fe, while W displays compact cascades and cascade overlap is not required to form large defects. Here, only an educated guess of the defect saturation level can be made. To estimate this level, as a consequence of displacement cascades, it is assumed that during the thermal spike that accompanies the cascade, within the region where the melting temperature is exceeded, the pre-existing displacement defects are erased. Gao and Bacon found that the radius of the molten zone $R_{melt}$ is related to the cascade energy $E_p$ by $R_{melt} \approx 3 a_0 E_p^{1/3}$, based on MD-results \cite{Gao1995}. Furthermore, using the NRT-model \cite{Norgett1975}, the amount of Frenkel pairs $N_F$ created during the same cascade of energy $E_p$ is approximately given by $N_F=0.4 \frac{E_{p}}{E_{d}}$, where $E_d=128$ eV is the displacement threshold energy for tungsten \cite{Setyawan2015}. Assuming a spherically shaped thermal spike, this gives $f_{max} \approx N_F/N_{melt} = 0.01$, where $N_{melt}$ is the number of atoms in the region that exceeds the melting temperature during the thermal spike. This is in the same range as the saturated defect fraction of 0.004-0.008 that was recently reported in \cite{Nordlund2018}, based on fcc studies. A lower bound for $f_{max}=0.004$ is estimated based on the reported defect densities for neutron-irradiated tungsten in the fast neutron reactor JOYO at irradiation temperatures of 400-756 \textdegree C \cite{Tanno2007, Fukuda2012}. In this work, the estimated value of $f_{max}=0.01$ will be used in the simulations.

\subsection{Mean-field recrystallization model}
\begin{figure}[ht!]
\centering
\includegraphics[width=0.4 \linewidth]{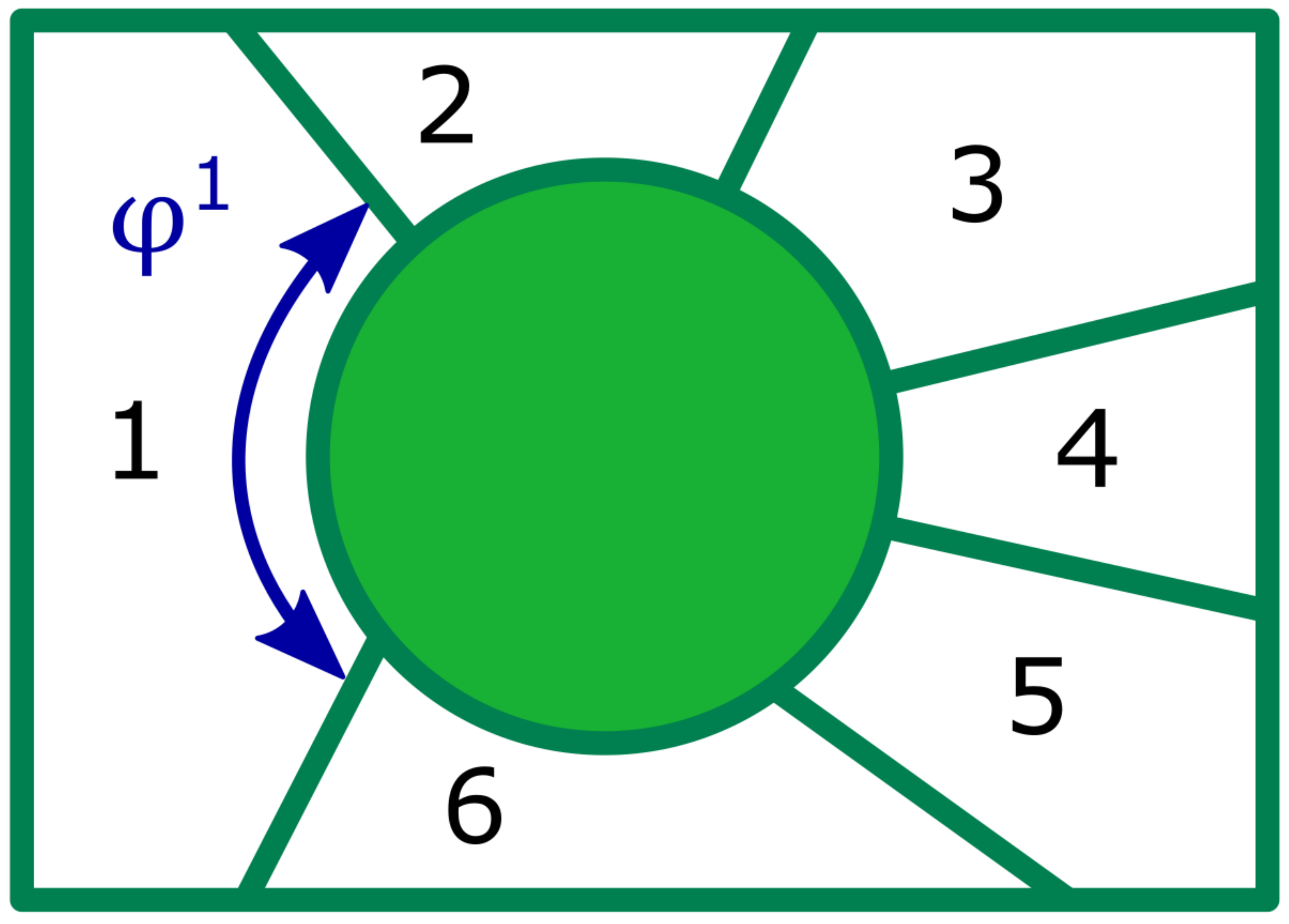}
 \caption{A representative grain, surrounded by 6 HEMs. The grain boundary surface fractions of the HEMs are denoted by $\phi$.}
 \label{fig:HEMs}
\end{figure}

The original multi-scale model \cite{Mannheim2018b} consists of only two homogeneous equivalent media (HEMs), following \cite{Bernard2011}. It is extended to comprise more than two HEMs, which enables an improved description of the grain growth behaviour. The microstructure consists of representative grains, which are considered spherical with properties: radius $r$, a number of grains $N$ that they represent, defect concentrations $C$ and dislocation density $\rho$. As illustrated in Figure~\ref{fig:HEMs}, each representative grain interacts with all the HEMs and each of those interactions results in the movement of a grain boundary segment. Together, the motion of the segments determines the amount of grain growth or shrinkage that occurs for a representative grain.  
The representative grains are assigned to a specific HEM based on their bulk stored energy densities, where for a number of HEMs $N_{HEM}$, there are $N_{HEM}-1$ energy limits; e.g. all grains for which $10^5$ J/m$^3 \le E^B < 10^6$ J/m$^3$ could be in HEM 2, all represenative grains for which $10^6$ J/m$^3 \le E^B < 2.5\e{6}$ J/m$^3$ in HEM 3, and so on. 

\paragraph{Grain growth} \mbox{}\\
Assuming that the grain boundary mobility $m$ is constant throughout the microstructure, the amount of grain growth depends on the stored energy difference $\Delta E$ between two neighbouring grains. The stored energy density $E_k$ of a spherical representative grain $k$ consists of a bulk contribution $E^B_k$ and a surface contribution $E^S_k$ and is given by
\begin{align} 
E_k = E^B_k + E^S_k = \sum_{n=1}^{N_{max}}\bigg[ C_{I_n,k}E^f_{I_n} + C_{V_n,k}E^f_{V_n}\bigg]+ \mu b^2 \rho_k/2  - 		TS_k + 3 \gamma_b /2 r_k,
\end{align}where $C_{I_{n,k}}$ is the concentration of interstitial clusters of size $n$ in grain $k$, $E^f_{I_n}$ is the formation energy for cluster $I_n$, and likewise for the vacancy clusters, $\mu$ is the shear modulus, $b$ is the magnitude of the Burgers vector, $\rho_k$ the dislocation network density in grain $k$, $S_k$ is the configurational (mixing) entropy, $\gamma_b$ the grain boundary energy and $r_k$ the radius of grain $k$. The adopted parameter values \cite{Li2012, Stoller1990,Lopez2015} can be found in Table~\ref{tab:parval}.
\\
In the mean-field model, grain growth is calculated based on the stored energy difference between a grain and the HEMs, $\Delta E = E^{HEM}_q - E_k$. Here, $E^{HEM}_q$ is the volume average of the stored energy density of all the grains that are contained in HEM $q$:
\begin{align}
E^{HEM}_q = \frac{\sum_{k \in q}r_k^3 N_k  E_k }{\sum_{\forall k} r_k^3 N_k}.
\end{align}The volume change of a grain $k$ that occurs in a time step $\Delta t = t_{i+1} - t_i$ is then given by: 
\begin{align}
\Delta V_k = \sum_{q=1}^{N_{HEM}} \Delta V^q_k = & \sum_{q=1}^{N_{HEM}} \phi^q 4 \pi r_k^2 v^q_k \Delta t = \sum_{q=1}^{N_{HEM}} \phi^q 4 \pi r_k^2 m (E^{HEM}_q - E_k) \Delta t, \label{eq:dv}
\end{align}where $\Delta V^{q}_k$ is the volume change of grain $k$ due to its interaction with HEM $q$, $N_{HEM}$ is the number of HEMs, $\phi^q$ is the fraction of the grain boundary surface area of a grain that is shared with HEM $q$, $v_k^q$ is the grain boundary velocity between HEM $q$ and grain $k$, $m$ is the grain boundary mobility. The grain boundary surface fraction that HEM $q$ shares with any representative grain is 
\begin{align}
\phi^q=\frac{A_q}{A_{tot}}=\frac{\sum_{k \in q}  N_k r_k^2}{\sum_{\forall k} N_k r_k^2}.
\end{align}The grain boundary mobility is taken to be \cite{Mannheim2018, Cram2009}:
\begin{equation}
m(T) 	= K_m \frac{ \beta \delta V_{m}}{b^2 RT} D^{GB}_0 \text{exp}\bigg( \frac{-Q^{GB}}{k_B T} \bigg),
\end{equation}where $K_m$ is characterized based on static recrystallization experiments on tungsten (\cite{Lopez2015}), $Q^{GB}$ is the activation energy for diffusion of tungsten along grain boundaries, $\delta$ is the thickness of the grain boundaries, $\beta$=0.3 is a fraction parameter \cite{Cram2009}, $V_{m}$ is the molar volume, $R$ is the gas constant and $D^{GB}_0$ is the self-diffusivity of tungsten along the grain boundaries. The parameters are specified in Table~\ref{tab:parval} in Appendix~\ref{app:cd}.\\
\\
In the model it is assumed that the moving grain boundaries sweep the defects they encounter. Therefore, defect-free volume is added to growing grains, which leads to a reduction in the average defect concentrations in the grain \cite{Bernard2011, Mannheim2018b}. The solution procedure for grain growth in this multi-HEM mean-field approach is detailed in Appendix~\ref{sec:sp}, where it is also explained how the grains are assigned to the HEMs, how the total amount of representative grains is bounded and how volume conservation is satisfied.

\paragraph{Nucleation} \mbox{}\\
Nucleation of new, defect-free, grains, is in principle expected to take place primarily at the grain boundaries for tungsten (necklace-type nucleation) \cite{Mannheim2018b, Lopez2015}. On top of that, under irradiation conditions, the displacement cascades may lead to a high lattice stored energy at the grain interior, which may trigger an additional nucleation mechanism: bulk nucleation. Both types of nucleation are illustrated in Figure~\ref{fig:scheme}. During nucleation, a small defected volume is replaced by a defect-free volume and new grain boundary surface area emerges. The process is driven by the reduction in the Gibbs free energy $\Delta E$, which is given by 
\begin{align}
\Delta E^S & = \frac{1}{K^S_a} \bigg[ - \frac{4 \pi r^3 }{3} ( E^{B} - E^B_0) + 3 \pi r^2 \gamma_b \bigg], \\
\Delta E^B & = \frac{1}{K^B_a} \bigg[ - \frac{4 \pi r^3 }{3} ( E^{B,HD} - E^B_0) + 4 \pi r^2 \gamma_b \bigg],
\end{align}for necklace and bulk nucleation respectively. Here, $r$ is the radius of the nucleus (new grain), $E^B=\sum_{q=1}^{N_{HEM}} \phi^q E^q$ is the average bulk stored energy density, $E^B_0$ is the bulk stored energy density for a grain with equilibrium defect concentrations, $E^{B,HD}$ is the average bulk energy density for all the grains with a (HD, high) bulk energy density that exceeds the nucleation threshold ($E^B_k > E^{B,thr}_{nuc}$)  and $K^S_a$ and $K^B_a$ are parameters for reduced activation energy during necklace and bulk nucleation respectively. A nucleation threshold applies, because nucleation is considered not to be possible at/in grains with a low defect density. Therefore, these low-defect density grains should not contribute to the average bulk stored energy, since this would lead to a higher nucleation activation barrier otherwise. \\
\\
The necklace and bulk nucleation rates $\dot{N}^S$ and $\dot{N}^B$ are given by: 
\begin{align}
\dot{N}^S & = K^S_N A_{nuc} \mathrm{exp} \bigg( \frac{-E^S_{act}}{k_BT}\bigg) \mathrm{exp} \bigg( \frac{-Q^{GB}}{k_BT} \bigg), \\
\dot{N}^B & = K^B_N V_{nuc} \mathrm{exp} \bigg( \frac{-E^B_{act}}{k_BT}\bigg) \mathrm{exp} \bigg( \frac{-Q^{GB}}{k_BT}\bigg).
\label{eq:nucrate}
\end{align}The nucleation rates depend on the nucleation activation energies $E^{S}_{act}$ and $E^B_{act}$, the temperature $T$, the nucleation surface area $A_{nuc}$, the nucleation volume $V_{nuc}$, the activation energy for the grain boundary mobility $Q^{GB}$ and the nucleation rate constants $K^S_N$ and $K^B_N$. Necklace nucleation takes place at all grain boundary segments, except for the segments where both the representative grain and the HEM have bulk energies below the nucleation threshold ($E^{B,thr}_{nuc}$).  This leads to the following expression for the nucleation surface area:
\begin{align}
A_{nuc} = 2 \pi \sum_i r_i^2 N_i \bigg[ 1 - \bigg( \frac{\sum_{j:E_j<E^{B,thr}_{nuc}} N_j r_j^2}{\sum_j N_j r_j^2 } \bigg)^2 \bigg],
\end{align}In this expression, the summations hold over all representative grains in the microstructure, unless specified otherwise. The nucleation volume $V_{nuc}$ for bulk nucleation consists of the volume of all the grains for which $E^B>E^{B,thr}_{nuc}$. Stable nuclei will form when $\frac{\partial \Delta E}{\partial t} < 0$ and when $v^{nuc}_{GB}>0$ (i.e. the nucleus is growing). Solving for these conditions gives the nucleation activation energies $E^S_{act}$ and $E^B_{act}$ and the nucleation radii $r^S_{nuc}$ and $r^B_{nuc}$, as detailed in Appendix~\ref{sec:sp}. 

\paragraph{HD daughter grains}\mbox{}\\
Grains that form by bulk nucleation are called HD daughter grains (HD = high defect density), and all the other grains are referred to as regular grains. The HD-daughter grains only grow with respect to the representative grains for which $E^B>E^{B,thr}_{nuc}$. HD-daughter grains are converted to regular grains as soon as their radius exceeds $r^{HD}/4$, where $r^{HD}$ is the average radius of all the grains for which $E^B>E^{B,thr}_{nuc}$, or when the bulk energy density of the HD-daughter grains exceeds $E^{B,thr}_{nuc}$ (see \cite{Mannheim2018b}). 

\subsection{Hardening}
The displacement defects in the lattice induce hardening of the material, as they form obstacles for dislocation motion. A hardness indicator $\mathcal{I}_H$, based on the Dispersed Barrier Hardening model, is used to qualitatively study the evolution of the hardness under irradiation \cite{Hu2016, Mannheim2018b}:
\begin{align}
\mathcal{I}_H = \frac{\sqrt{\rho}+\frac{M}{\alpha_T}\sqrt{\sum_j2\alpha_j^2C_jr_j}}{\sqrt{\rho_0}}.
\end{align}Here, $M$ is the Taylor factor, $\alpha_T$ is the dislocation barrier strength, $\rho_0$ is the initial dislocation density, $\alpha_n$ is the defect barrier strength of defect type $n$ (see Table~\ref{tab:dbh}) and $r_n$ is the radius of defect type $n$, see Equations~\ref{eq:In} and \ref{eq:Vn} in Appendix~\ref{app:rn}.

\begin{table*}[ht!]
\caption{Barrier strengths as used in the DBH-model, based on \cite{Hu2016}, for the cluster sizes as listed. For clusters of smaller sizes, $\alpha=0$ is taken.}
\begin{tabular}[ht!]{l | c | c  | c }
 			& Barrier strength factor $\alpha$ 	&     Diameter (nm) 		&  Cluster size $N$ 		\\ \hline
Interstitial loop       & 	0.15 					& 	1.0-2.7	 	& 14-100 			\\
Void 			& 	0.25 					& 	1.0-1.4 		& 33-100 			
\label{tab:dbh}
\end{tabular}
\end{table*}

\subsection{Parameter characterization} \label{sec:parkar}
\begin{figure}[ht!]
\subfloat[]{\includegraphics[height=0.22\textheight]{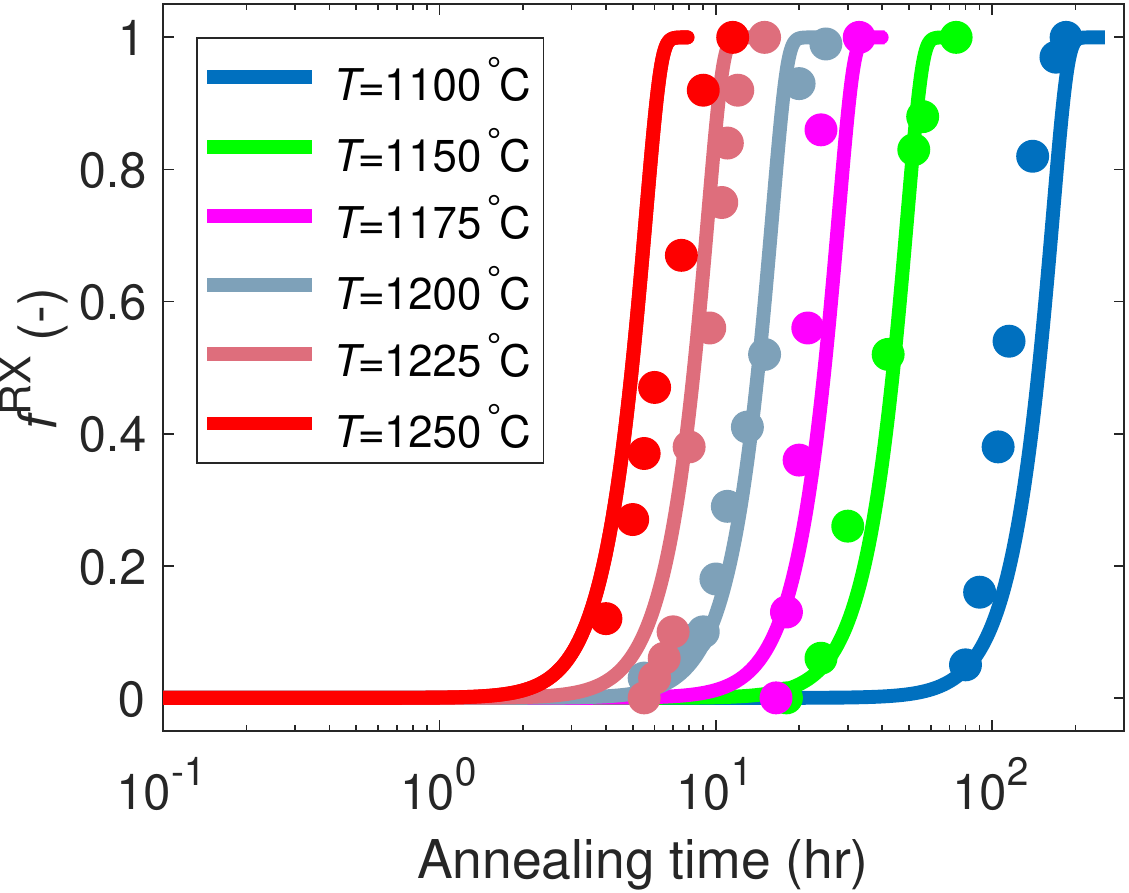}} \hspace{0.2cm}
\subfloat[]{\includegraphics[height=0.22\textheight]{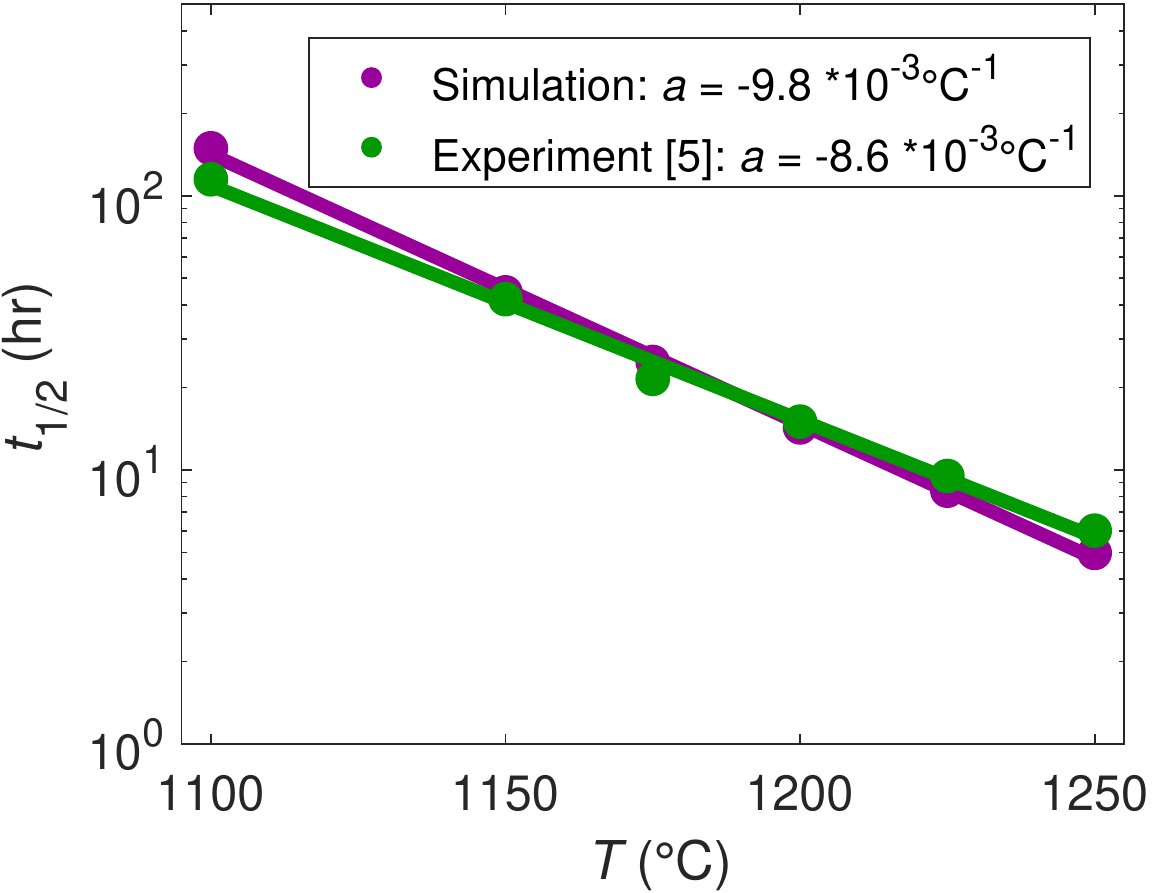}} \vspace{0.01cm}
\subfloat[]{\hspace{0.15cm} \includegraphics[height=0.22\textheight]{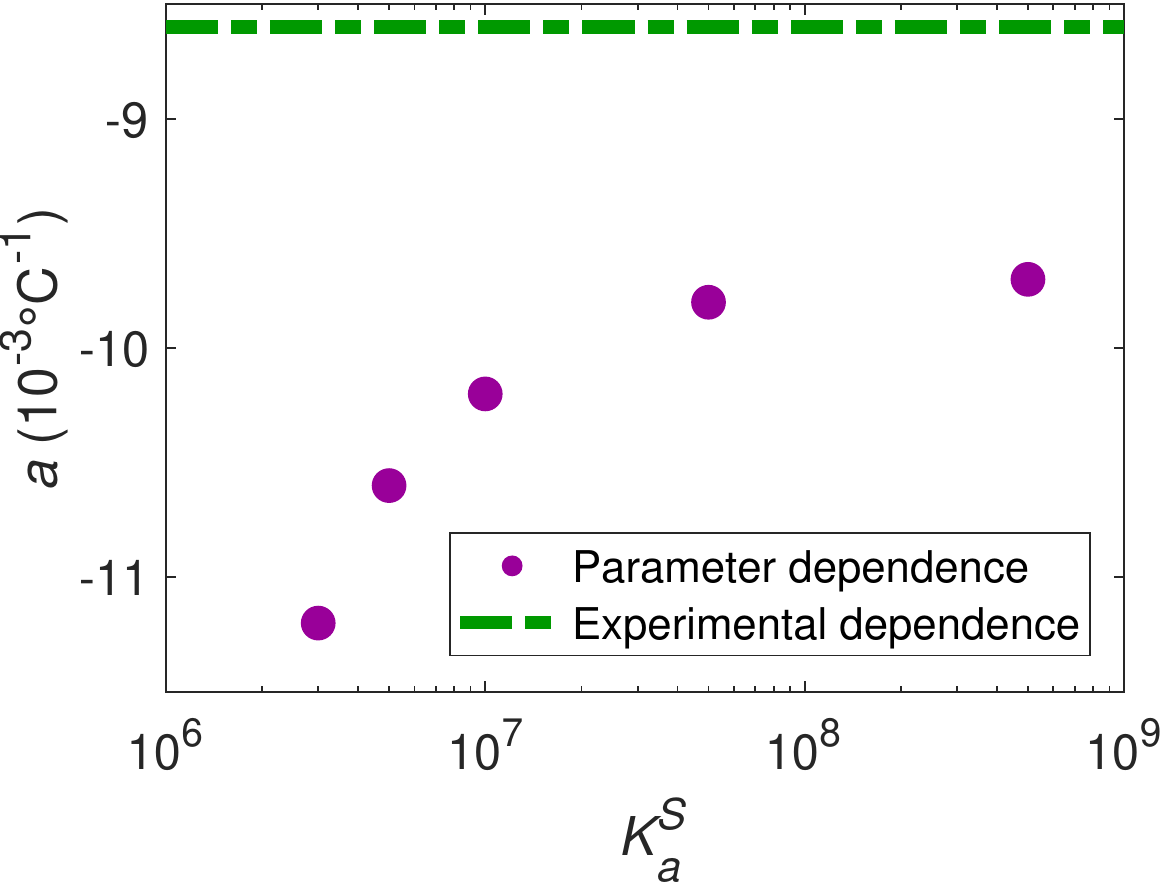}}
\caption{(a) Simulation results for the evolution of the recrystallized volume fraction $f^{RX}$ as a function of annealing time during static recrystallization, for various temperatures, compared to experimental results obtained by Lopez \cite{Lopez2015}, from hardness measurements and EBSD, for warm-rolled samples that were thickness-reduced with 90\%. (b) Temperature-dependence of the half-time for recrystallization, including a fit of the slope, for the experimental results and for the simulations; (c) Influence of $K^S_a$ on the temperature-dependence of the speed of recrystallization, slope $a$.}
\label{fig:parchar}
\end{figure}

The parameters $K_m$, $K_N$ and $K_a$, that are related to the grain boundary mobility, the nucleation rate and the reduced nucleation activation energy, have been identified by performing static recrystallization simulations using the mean-field recrystallization model and the experimental results of Lopez \textit{et al.} \cite{Lopez2015}. Reparameterization (with respect to \cite{Mannheim2018}) is performed here because in the improved model, the surface fractions are calculated differently. Only necklace nucleation is assumed to take place here, based on experimental observations of static recrystallization of tungsten \cite{Lopez2015}. The initial and final microstructural parameters are listed in Table~\ref{tab:micr}. The original microstructure consisted of 500 representative grains with normally distributed sizes and dislocation densities. 2 HEMs were used, with the bulk energy threshold for the second HEM taken as 10$^6$ J/m$^3$ and with a nucleation energy density threshold of $E^{B,thr}_{nuc}=10^6$ J/m$^3$.
\begin{table}[ht!]
    \begin{tabular}{ p{6 cm} |  p{2.5 cm} | p{1.5 cm} }

	\textbf{Property} & \textbf{Value} & \textbf{Source}\\ \hline
	& \\
	\textit{Initial microstructure} & \\ 
	Initial average grain size $\bar{r}_i$ & 18.6 $\mu$m & \cite{Lopez2015}  \\
	Standard deviation for $\bar{r}_i$ & 3.1 $\mu$m \\ 
	Initial average dislocation density $\bar{\rho}_i$ & 3.2\e{14} m$^{-2}$ & \cite{Lopez2015} \\
	Standard deviation for $\bar{\rho}_i$ & 5.2\e{13} m$^{-2} $ \\ 
	& \\
	\textit{Final microstructure} & \\
	Final average grain size $\bar{r}_f$  & 54.1 $\mu$m & \cite{Lopez2015}\\ 
	(after 25 hr at \textit{T}=1200 $\degree$C) & \\
	Final dislocation density $\rho_f$ & 1\e{9} m$^{-2} $ & \cite{Li2012}\\ \hline
	\end{tabular}
	\caption{Initial and final microstructural properties, before and after static recrystallization, based on \cite{Lopez2015}.}
	\label{tab:micr}
\end{table}The obtained fit (for $K^S_a$=$5\e{7}$, $K_m$=1490 and $K^S_N=2.5\e{17}$m$^{-2}$s$^{-1}$) is shown in Figure~\ref{fig:parchar}a and is in adequate agreement with the experimental data. Initially, the parameter identification was done for $T$=1200 $\degree$C; once a satisfactory fit for this temperature was found, the temperature dependence was also taken into the consideration. Parameters $K^S_a$ and $K_m$ and $K^S_N$ were determined such that full recrystallization would be obtained after 25 hours, with a final grain size of 54.1 $\mu$m. It was found that given $K^S_a$, a unique set of $K_m$ and $K^S_N$ exists that fullfills these requirements, as illustrated in Appendix~\ref{app:srx}. 
Figure~\ref{fig:parchar}b shows the temperature dependence of the time required to reach a recrystallized volume fraction $f^{RX}$=0.50, both for the experimental results and for the simulation results, expressed by the slope $a=\Delta log_{10}(t_{1/2})/\Delta T$ of the recrystallization time as a function of temperature. The slope was determined for various choices of $K^S_a$ (with the corresponding $K_m$ and $K_N$ such that the expected final grain size and expected recrystallization time is reached at 1200 \textdegree C). It was found that the slope $a$ depends on the parameter $K^S_a$, as shown in Figure~\ref{fig:parchar}c. This dependence converges to a plateau and it is not possible to choose $K^S_a$ such that the experimental slope is recovered. Below, the influence of the choice of $K^S_a$ on the microstructural evolution under neutron irradiation will be shown. For each of the choices for $K^S_a$ (with the corresponding $K_m$ and $K^S_N$), the shape of the evolution of the recrystallized volume fraction $f^{RX}$ under static recrystallization is the same (not shown). However, there is another difference: a lower value for $K_a$ is accompanied by a higher value of $K^S_N$ and a nucleation rate that is initially higher and decreases afterwards, while a high value for $K^S_a$ leads to lower nucleation barrier and a nearly constant nucleation rate. The total amount of new grains over the course of the simulation is the same for both cases. The final grain size distribution will be slightly broader in the case of a high value for $K^S_a$, because of the constant nucleation rate, that leads to more heterogeneous sizes for the new grains.
\begin{figure}[ht!]
\subfloat[]{\includegraphics[height=0.22\textheight]{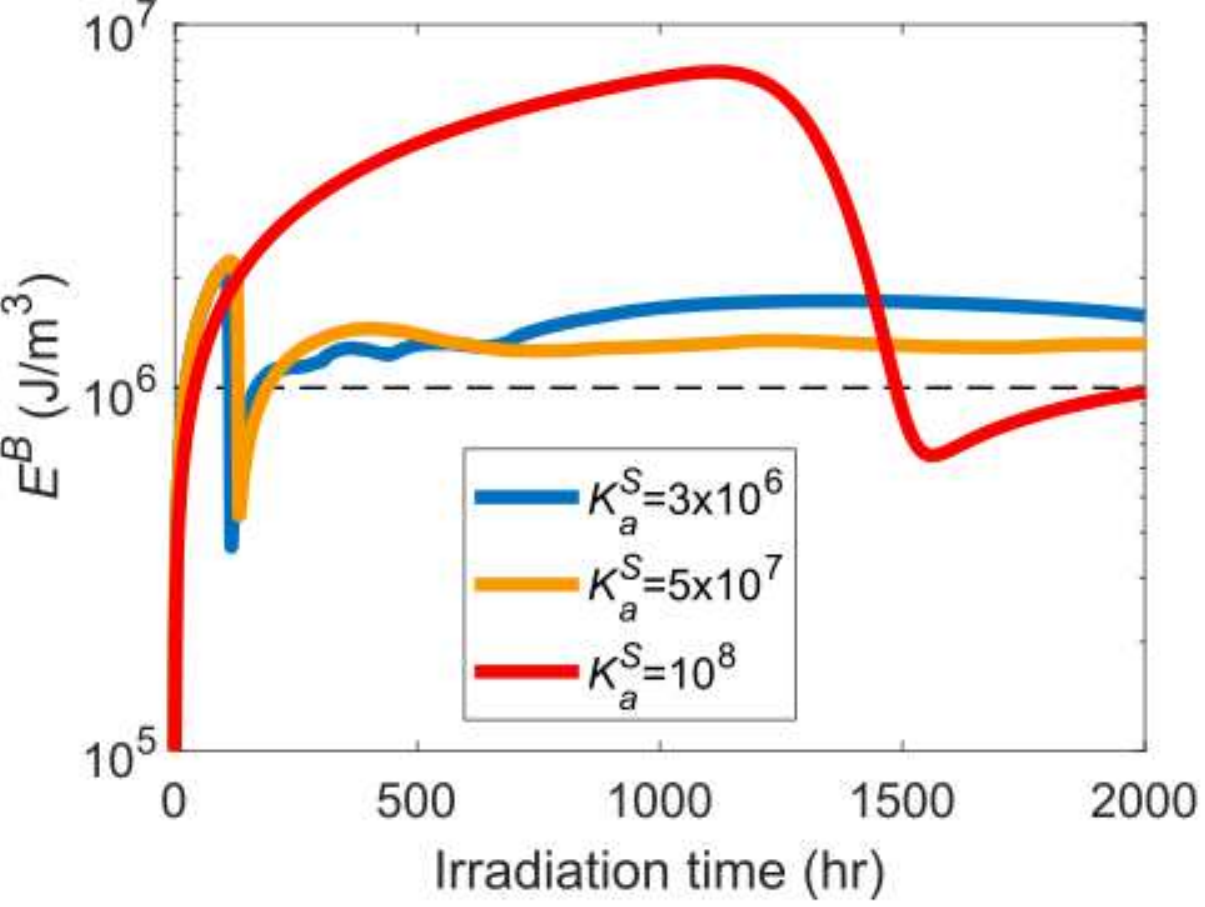}} \vspace{0.1cm}
\subfloat[]{\includegraphics[height=0.22\textheight]{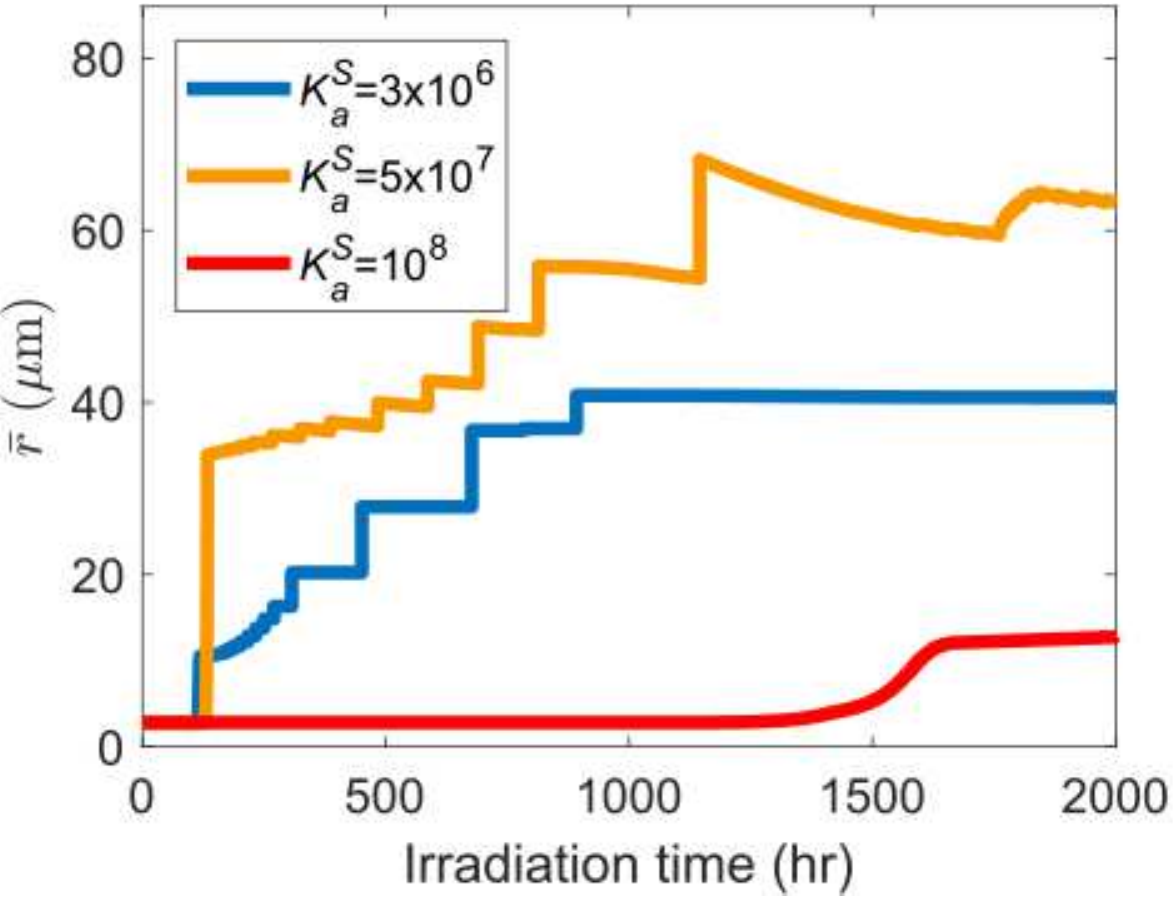}}
\caption{(a) Evolution of the bulk stored energy density $E^B$ and of the (b) average grain size $\bar{r}$ for different parameter sets during neutron iradiation at $T$=1100 \textdegree C. The dashed line shows the threshold that separates the two HEMs for the simulation with $K^S_a=10^8$.}
\label{fig:pardep}
\end{figure}

\begin{table}[ht!]
	\begin{tabular}{| p{1.8 cm} |  p{2.5 cm} | p{2 cm}| p{1.5 cm} | p{2.5 cm} |}\hline
	$K^S_a$ 	&  $K^S_N$ (m$^{-2}$s$^{-1}$)					& $K_m$ 	& $K_{m_0}$   & Previous/current	\\ \hline
	3\e{6} 	&  $9.3\e{26}$ 							& 1335 	& 	- 		& Current \\ \hline
	5\e{7} 	&   2.5\e{17}							& 1490  	& 	- 		& Current \\ \hline
	1\e{8}  	& 3.16\e{17} 							& 1.8\e{4} 	& 	25 	& Previous 	\\ \hline 
	\end{tabular}
	\caption{Two currently possible parameter sets identified from the experimental data and the previous parameter set, that includes a reduced grain boundary mobility for pinned grains, $K_{m_0}$.}
	\label{tab:parsets}
\end{table} 

\paragraph{Parameter effects on the full model}\mbox{}\\
Based on the results shown in Figure~\ref{fig:parchar}, there is too little information to determine for which value of $K^S_a$ the reality is best represented, other than that for a higher value of $K^S_a$, the temperature dependence is slightly better matched. In Figure~\ref{fig:pardep}a and b it is shown how much the results for the full model for neutron-induced recrystallization are affected by the set of recrystallization parameters. Simulations were performed using $K^S_a=3\e{6}$ and $K^S_a=5\e{7}$ (see Table~\ref{tab:parsets}) at $T$=1100 \textdegree C and are compared to the simulation results that were obtained with $K^S_a=1\e{8}$ (see Table~\ref{tab:parsets}), using the previous method for calculating the surface fractions and using only 2 HEMs. For the present simulations 20 HEMs were used, with 50 representative grains in each HEM. From Figure~\ref{fig:pardep}a and b it becomes clear that the  recrystallization parameters clearly affect the evolution of the mean grain size, whereas the bulk stored energy evolves similarly for either of the two recrystallization parameter sets. These results show thatin spite of the fact that the values for the nucleation parameters are uncertain, the multi-scale model is not very sensitive to the choice of these parameters. To obtain more accurate recrystallization parameters, the model could be parameterized better with experimental data that entail the final grain size and halftime $t_{1/2}$ of recrystallization for several temperatures. Additional information about the evolution of the grain size distribution would improve the parameterization as well. \\
\\
The experimentally determined temperature-dependency of the recrystallization half-time, captured with slope $a$, is uncertain for several reasons: (1) at high annealing temperatures, when the recrystallization half-time is small, the effect of the ramp time to heat the sample may be non-negligible; (2) for most temperatures, the recrystallized fraction was only determined from hardness measurements, whereas EBSD-data would be more reliable and (3) the information on the initial and final grain size is not very precise. Nevertheless, to capture the current experimental temperature-dependency of the recrystallization half-times correctly, further extensions of the mean-field model for recrystallization would be needed. For example, it may then be necessary to include the effect of the grain shapes and/or grain orientations on the grain boundary mobility.\\
\\
In the remainder of this article, the parameter set with $K_a$=5\e{7} will be used. 

\section{Results}

\subsection{Isothermal microstructural evolution}
\begin{figure}[H]
\subfloat[]{\includegraphics[height=0.22\textheight]{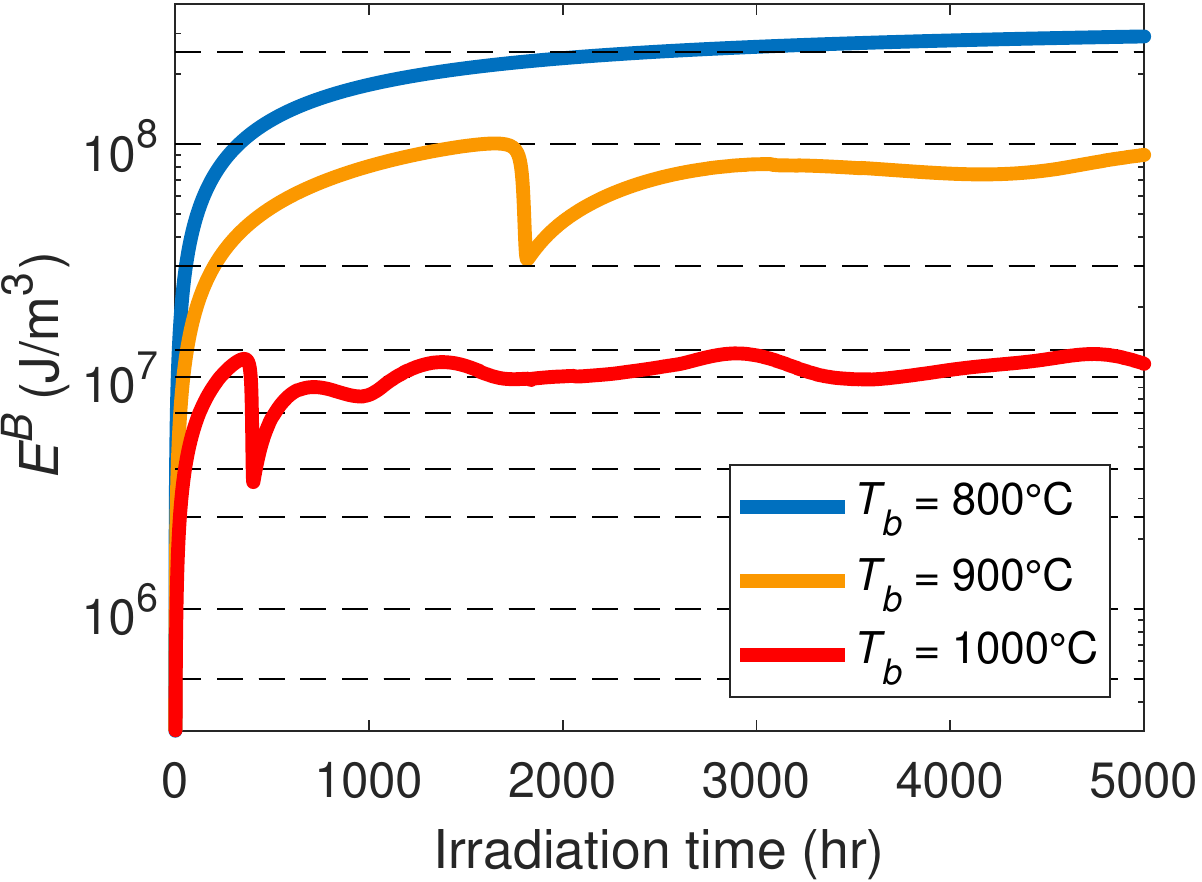}}
\subfloat[]{ \includegraphics[height=0.22\textheight]{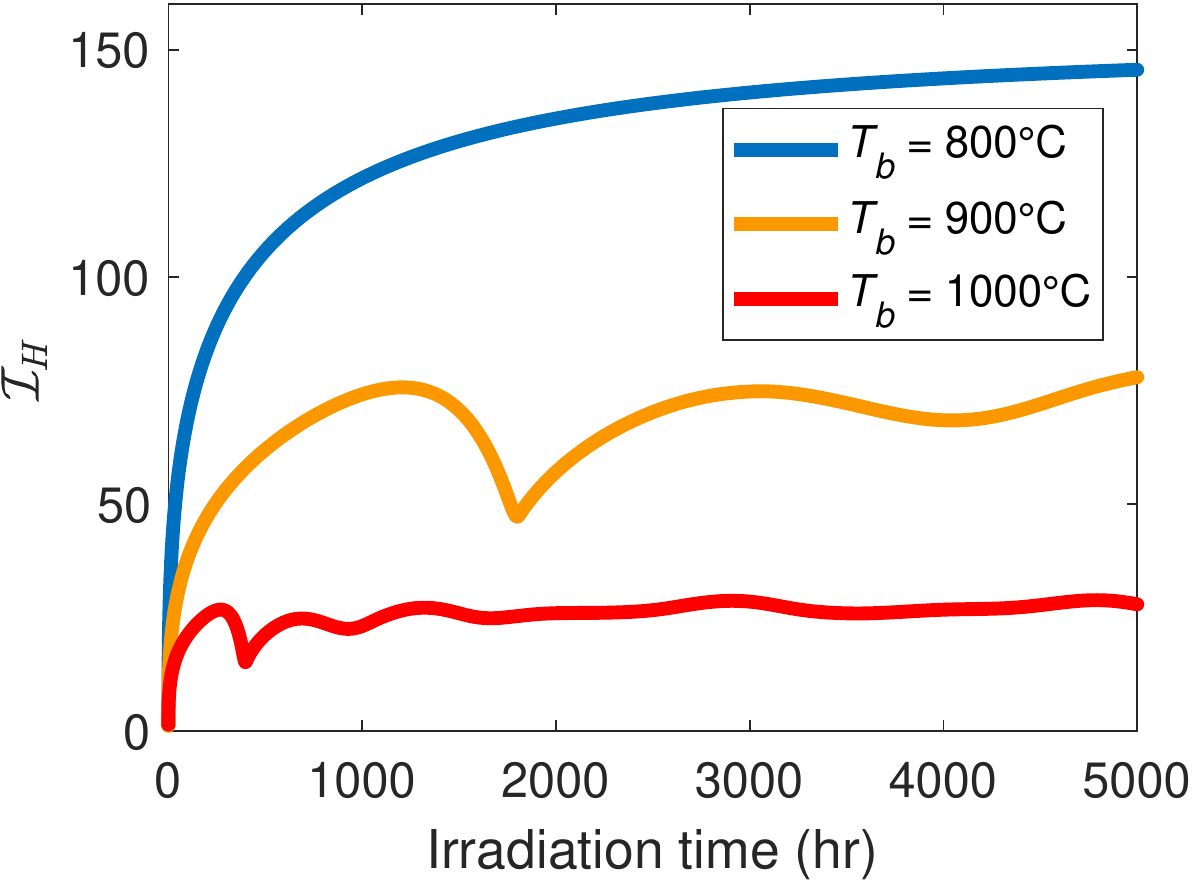}} \vspace{0.01cm}
\subfloat[]{\includegraphics[height=0.22\textheight]{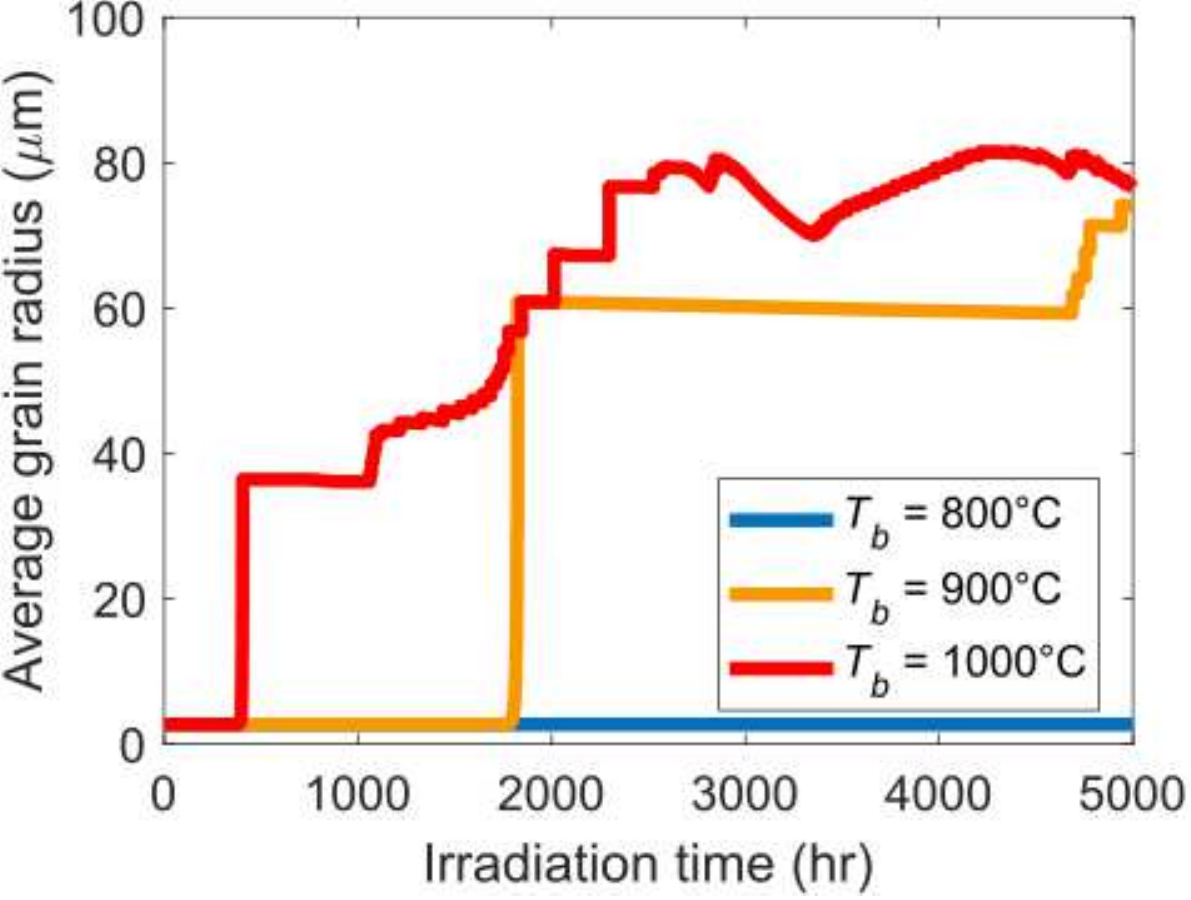}}
\subfloat[]{\includegraphics[height=0.22\textheight]{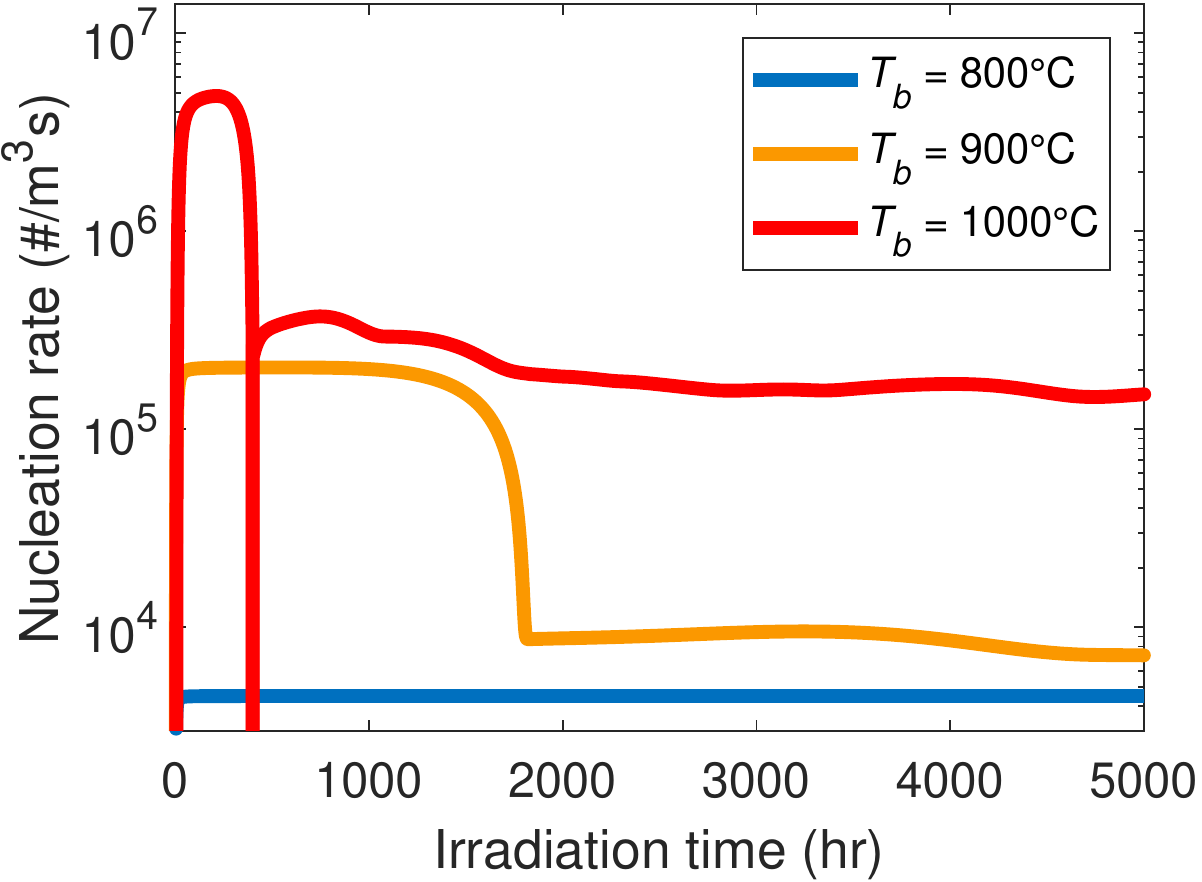}}
\caption{Temporal evolution of (a) the bulk stored energy density $E^B$, (b) the hardness indicator $\mathcal{I}_H$, (c) the average grain radius $\bar{r}$ and (d) the nucleation rate, during irradiation at temperatures 800 \textdegree C, 900 \textdegree C and 1000 \textdegree C. The dashed lines in (a) indicate the energy limits of the HEMs. }
\label{fig:isolow}
\end{figure}

Figure~\ref{fig:isolow} shows the microstructural evolution, as expressed by the bulk energy density $E^B$, the average grain size $\bar{r}$, the hardness indicator $\mathcal{I}_H$ and the volume fraction of original grains during irradiation at temperatures of 800\textdegree C, 900\textdegree C and 1000\textdegree C, considering necklace nucleation only. The simulation settings for the representative grains and the HEMs can be found in Appendix~\ref{app:cd}. Note that $E^B$ is defined as the summation of the energies of the HEMs, weighted with their respective surface fractions, while for $\mathcal{I}_H$, the overall volume averaged concentrations are used.

For the adopted nucleation parameters, recrystallization takes place rapidly at an irradiation temperature of 1000 \textdegree C. The hardness indicator stays under the value of 30. For 900 \textdegree C, recrystallization only sets in after more than 1000 hours of irradiation with a maximum value for the hardness indicator of 80. For 800 \textdegree C, no recrystallization takes place in the first 4000 hrs and the hardness indicator approaches 150 and keeps increasing. Figure~\ref{fig:isolow} reveals that cyclic recrystallization takes place for the temperatures of 900 \textdegree C and 1000 \textdegree C, indicating that the average grain size keeps increasing until it saturates (as a result of the balance between the nucleation rate, the amount of damage by irradiation and the amount of recovery by growth of the nucleated grains). A potential reduction of the high hardness found at 800 \textdegree C, using heat treatments, is thereby suggested. Therefore, 800 \textdegree C is selected as the base temperature for simulations of non-isothermal heat treatments. 

\subsection{Non-isothermal microstructural evolution}
\begin{figure}[H]
\subfloat[]{\includegraphics[height=0.22\textheight]{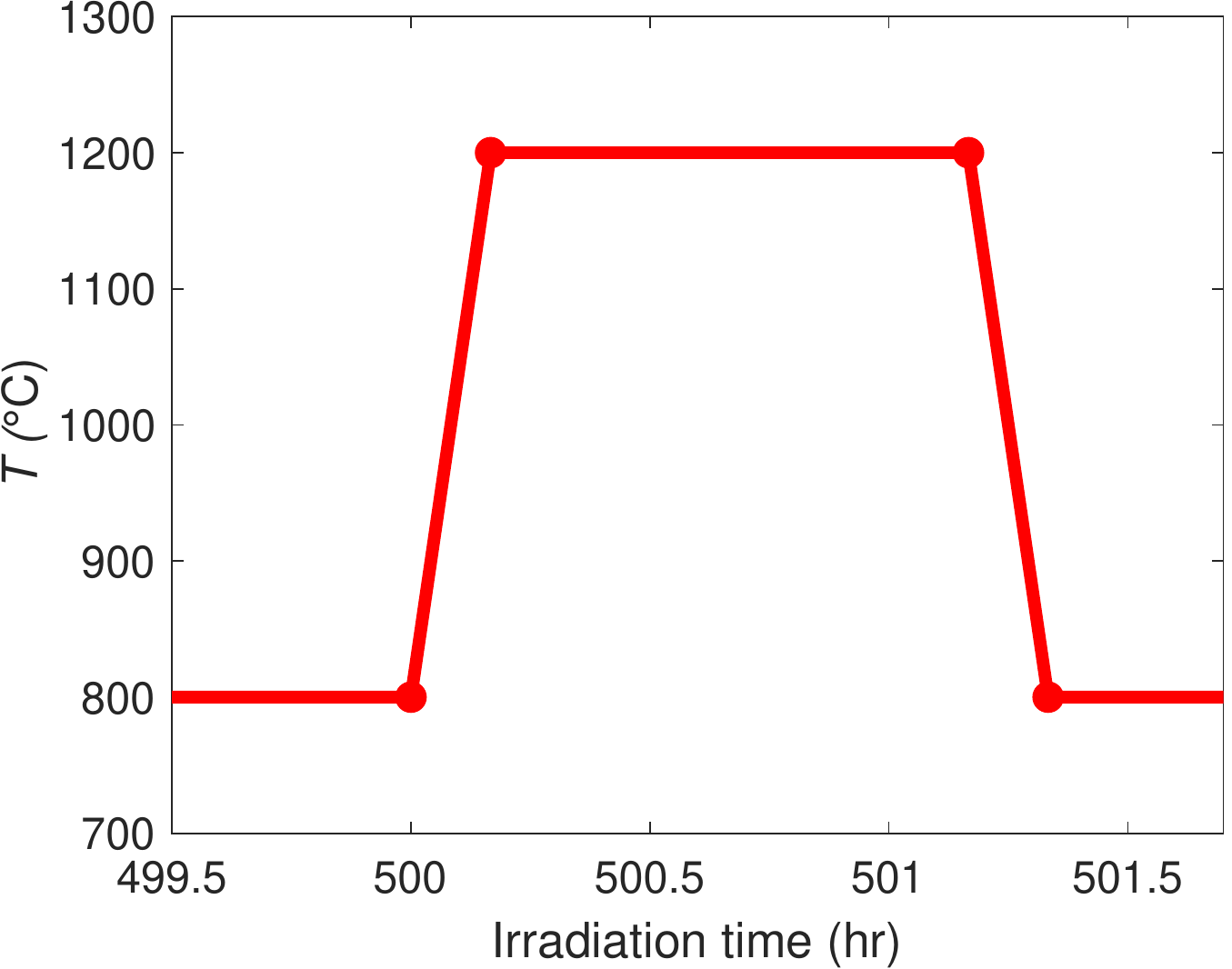}} \hspace{0.1cm}
\subfloat[]{\includegraphics[height=0.22\textheight]{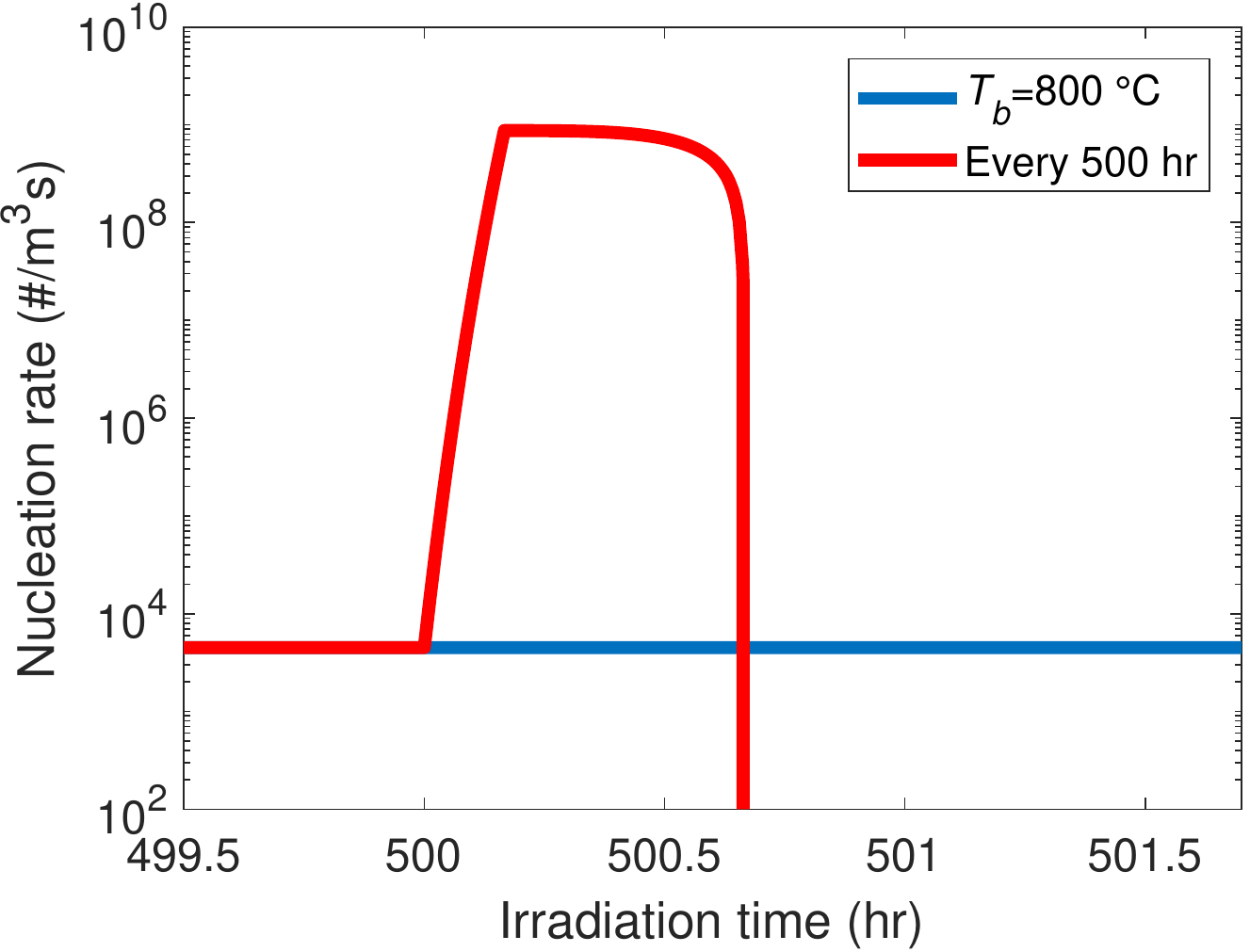}} \vspace{0.01cm}
\subfloat[]{ \includegraphics[height=0.22\textheight]{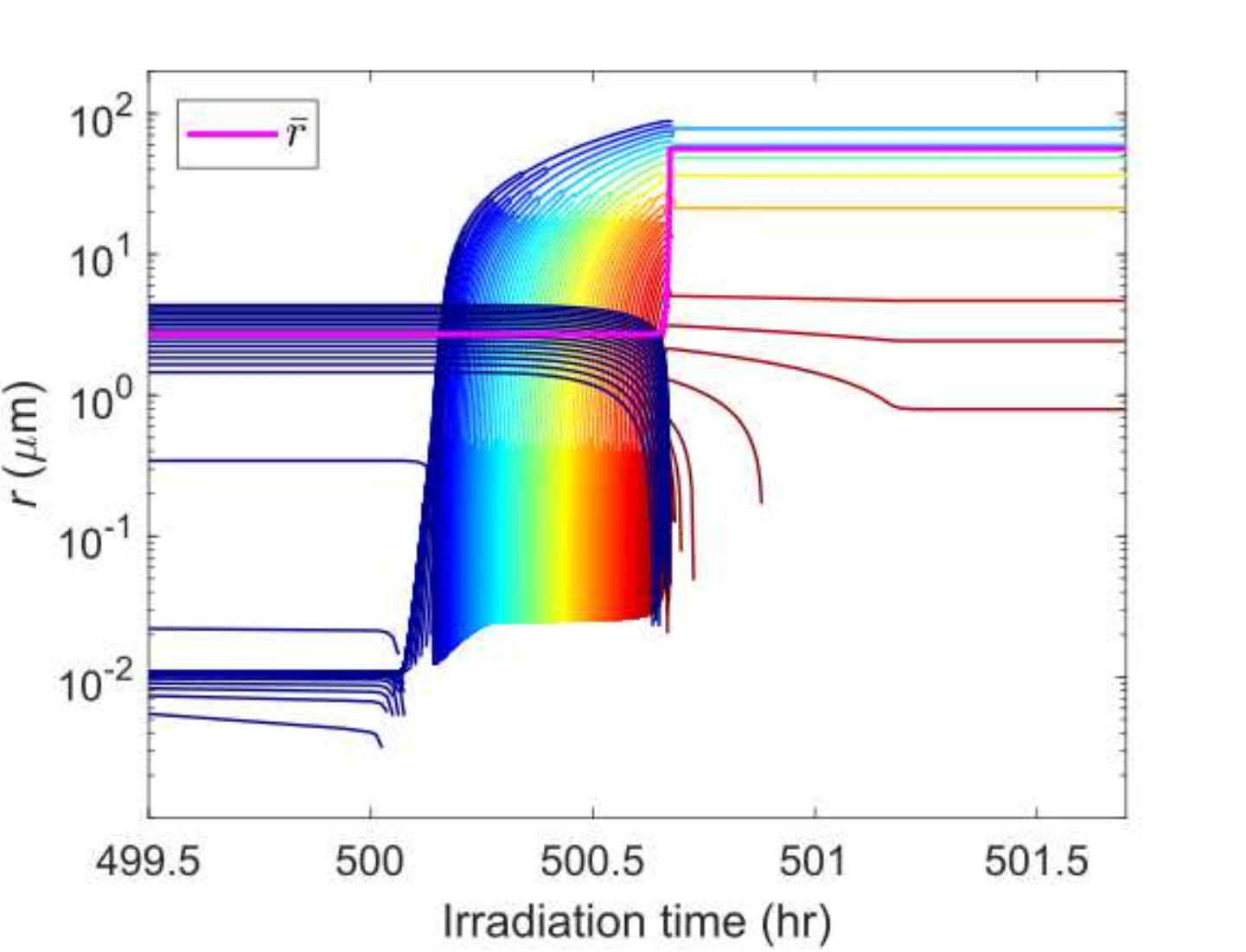}}\hspace{0.1cm}
\subfloat[]{\includegraphics[height=0.22\textheight]{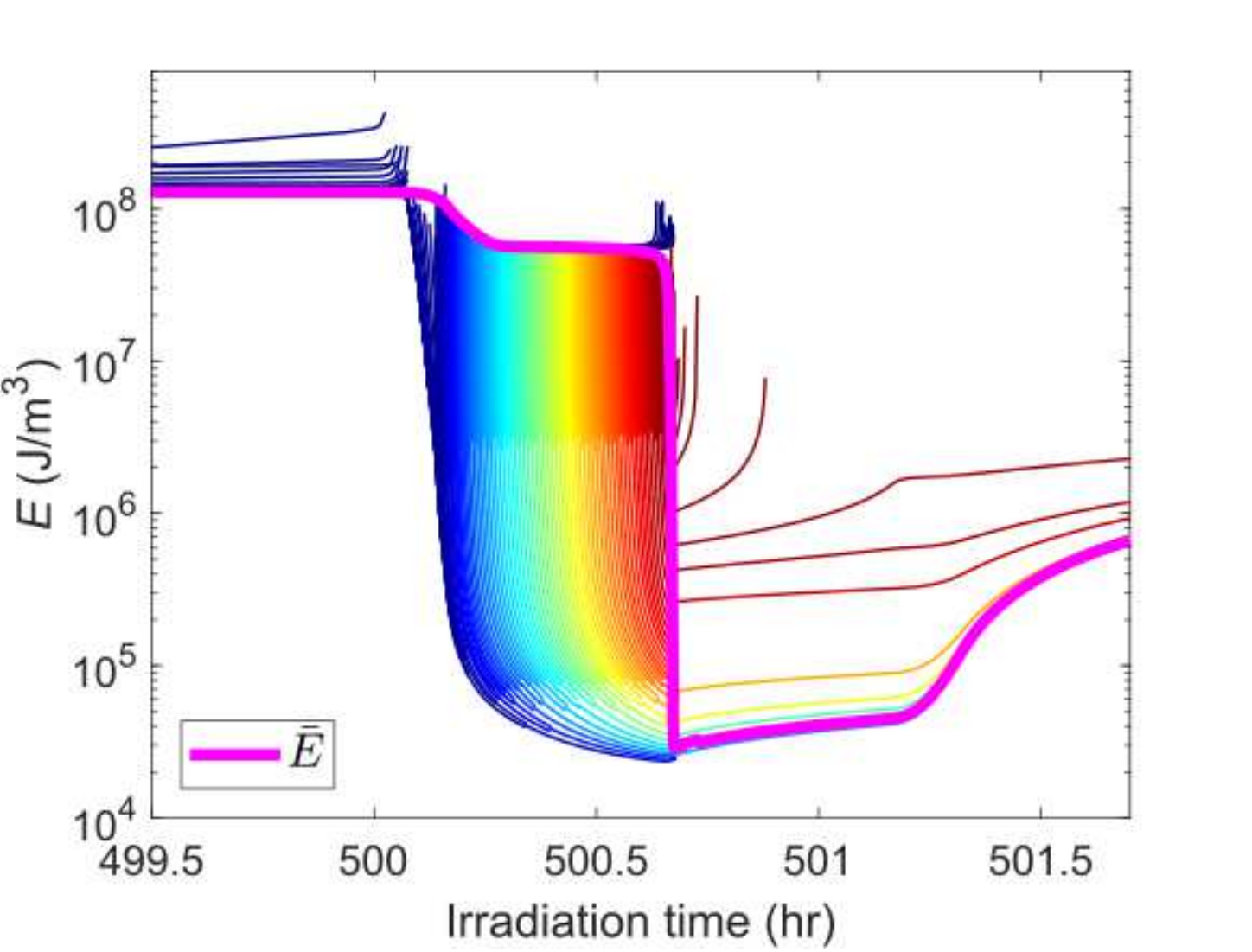}}
\caption{(a) Typical temperature profile during the applied heat treatment; (b)-(d) Microstructural evolution during heat treatment;(b) Nucleation rate (in comparison to the nucleation rate in the isothermal simulation at 800 \textdegree C); (c) Evolution of the grain radius and (d) evolution of the stored energy density of the representative grains. The colors in (c-d), from blue to red, indicate the order in which the new grains formed. Lines end when grains are completely consumed by other grains, or when grains merge with other representative grains, which occurs when the maximum amount of representative grains in a HEM is exceeded, see Appendix~\ref{sec:sp}.}
\label{fig:profile}
\end{figure}

\begin{figure}[H]
\subfloat[]{\includegraphics[height=0.22\textheight]{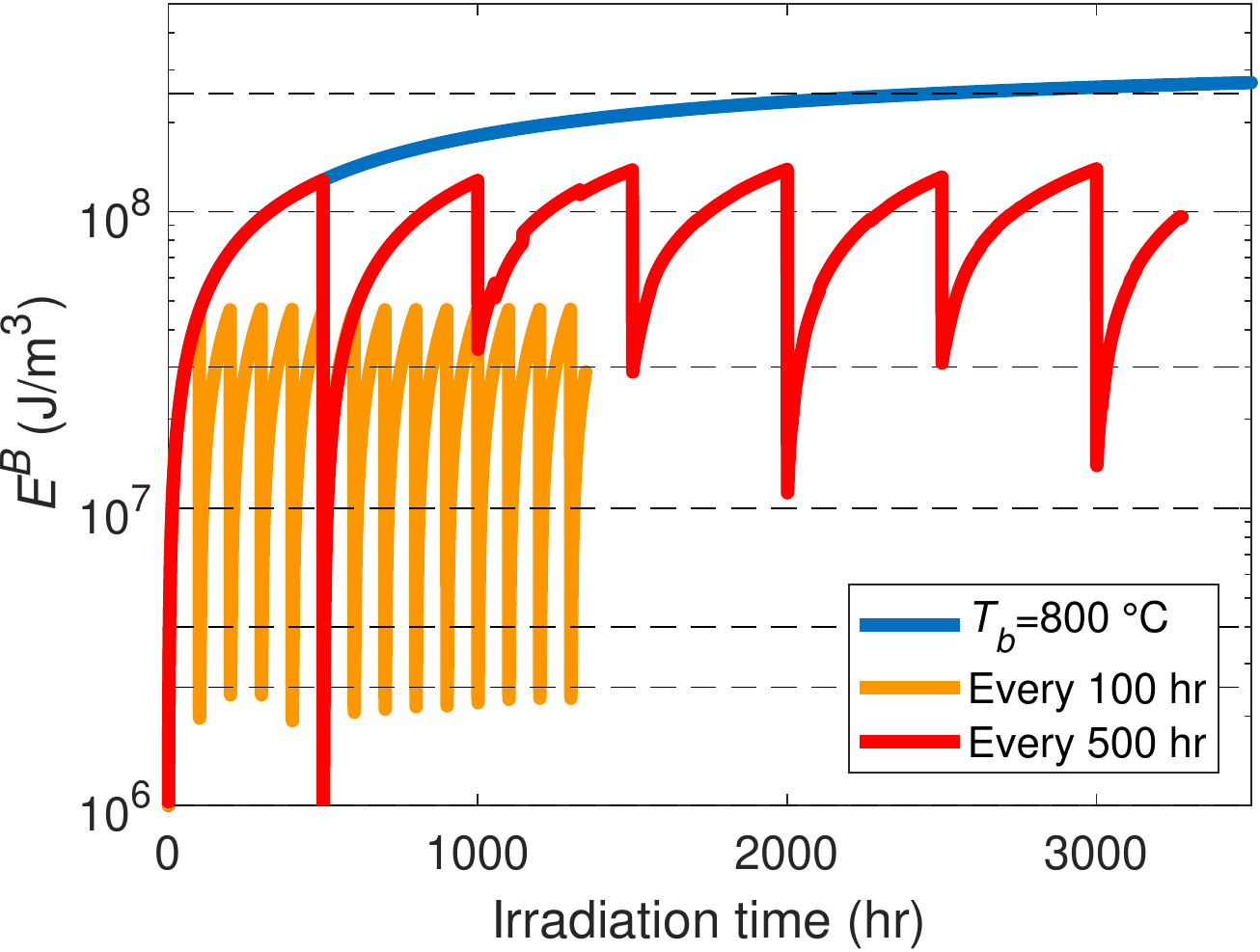}}
\subfloat[]{ \includegraphics[height=0.22\textheight]{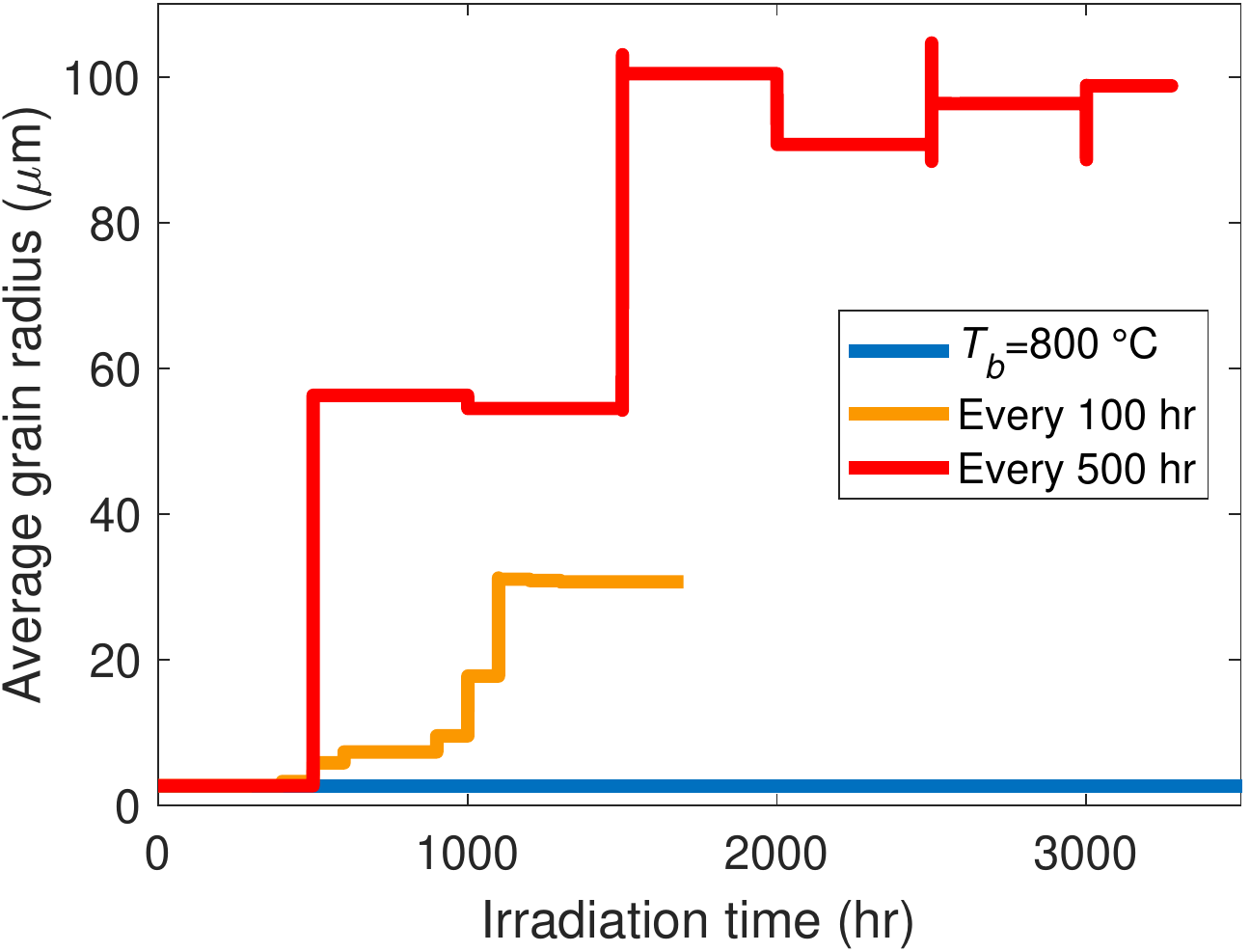}} \vspace{0.01cm}
\subfloat[]{\includegraphics[height=0.22\textheight]{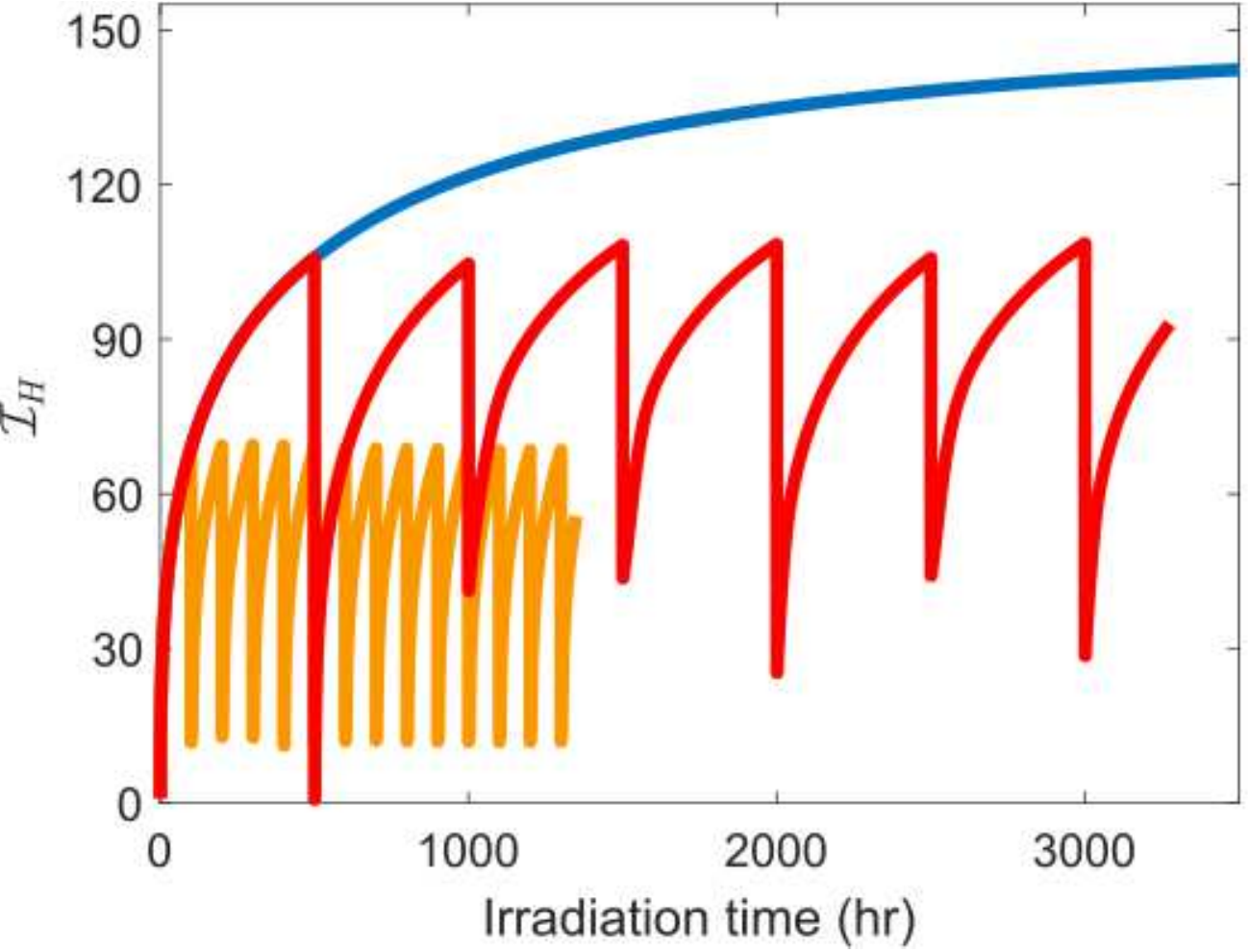}}
\subfloat[]{\includegraphics[height=0.22\textheight]{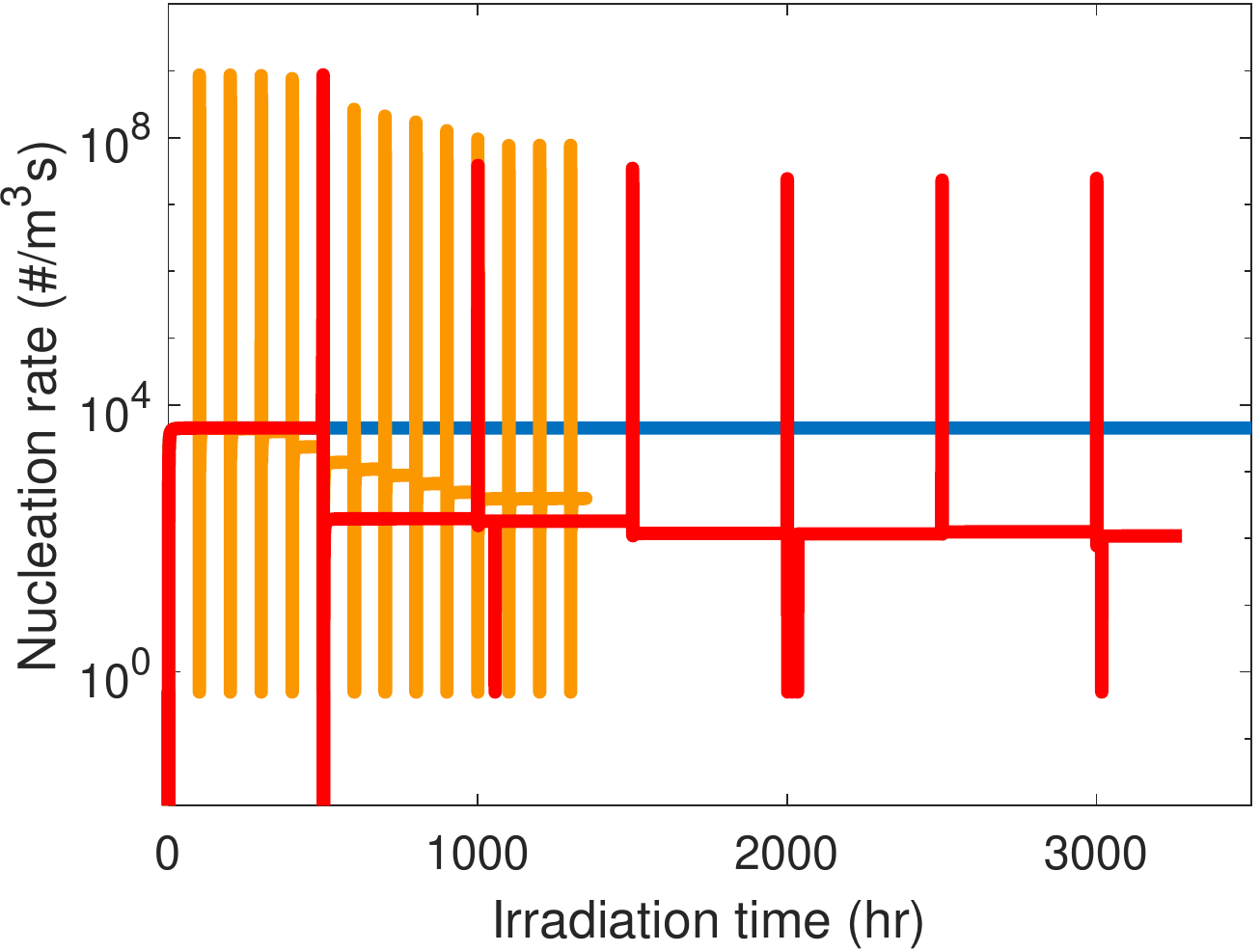}}
\caption{Temporal evolution of (a) the bulk stored energy density $E^B$, (b) the hardness indicator $\mathcal{I}_H$, (c) the average grain radius $\bar{r}$ and (d) the nucleation rate, during irradiation, for repeated heat treatments every 100 hr, every 500 hr, in comparison to the evolution under isothermal conditions. The dashed lines in (a) indicate the energy limits of the HEMs.}
\label{fig:noniso}
\end{figure}

In the next simulations, it will be explored how much recovery can be obtained by heating repeatedly for a short amount of time. Figure~\ref{fig:profile}a shows the typical temperature profile that is applied every 500 hrs (or every 100 hrs) in the next simulations. The material is heated in 10 min to the annealing temperature of 1200 \textdegree C, at which it is kept for 1 hr and after which the material is cooled back to 800 \textdegree C in 10 min. At first, only necklace nucleation is considered, and thereafter, the combination of necklace and bulk nucleation is considered.
Figures~\ref{fig:profile} b-d show the microstructural evolution that occurs during the first heat treatment cycle (annealing at 1200 \textdegree C) after 500 hrs of irradiation at 800\textdegree C: the nucleation rate rises (Figure~\ref{fig:profile}b) because of the increased temperature, which affects the exponential terms in Equation \ref{eq:nucrate}. The increased grain boundary mobility and the high amount of new nuclei lead to a fast, complete recrystallization: the original representative grains shrink in size until they vanish (Figure~\ref{fig:profile}c) and the stored energy density of the nucleated grains (Figure~\ref{fig:profile}d) decreases, while they grow to become large grains. Upon full recrystallization, the average stored energy density (pink line) drops and the average grain radius increases, as shown in Figures~\ref{fig:profile}c and d. Once the average stored energy density decreases, the nucleation rate also drops. The material is cooled down to 800 \textdegree C again, and up to 1000 hrs (when the next cycle of extra heating starts) the mobility of the grain boundaries and point defects again decreases, making recovery difficult, whereas the stored energy due to lattice damage will accumulate again.\\
\\
The effects of repeated heating, every 100 hrs and every 500 hrs, are shown in Figure~\ref{fig:noniso} and are compared to the isothermal irradiation situation. 
In the isothermal situation, the hardness  indicator reaches 140, see Figure~\ref{fig:noniso}c. By heating every 500 hrs for 1 hr to 1200 \textdegree C,  its value can be limited to a value below 110 and by heating more frequently, every 100 hrs, it stays below 70. For the 500 hrs repetitive heat treatment, for the first cycle, full recrystallization takes place, where $E^B$ drops by several orders of magnitude (Figure~\ref{fig:noniso}a). During the second cycle, the amount of grain boundary area per volume is smaller, as the average grain size is larger (Figure~\ref{fig:noniso}b). This leads to a lower nucleation rate (Figure~\ref{fig:noniso}d) and full recrystallization is no longer achieved ($E^B$ does not drop below $5 \times 10^7$ J/m$^3$ at 1000 hrs). Nevertheless, $\mathcal{I}_H$ reduces considerably and it does not seem  to be critical to reach full recrystallization, as long as a sufficient amount of recovery is achieved. If the annealing is applied every 100 hrs, full recrystallization is not achieved during the first cycle, as the driving force for recrystallization is lower after 100 hrs of irradiation, which means that the full recrystallization process completes less easily within 1 hr of annealing at 1200 \textdegree C. The expected maximum increase of the hardness indicator for a certain interval of time between cyclic heat treatments is related to its evolution under isothermal conditions (Figure~\ref{fig:noniso}c): each new cycle following a heat treatment entails a hardness indicator that increases to the same level at the start of the first cycle. Yet, the peaks in $\mathcal{I}_H$   can be slighly higher if the recrystallization was incomplete during the previous cycle.

\begin{figure}[H]
\subfloat[]{\includegraphics[height=0.20\textheight]{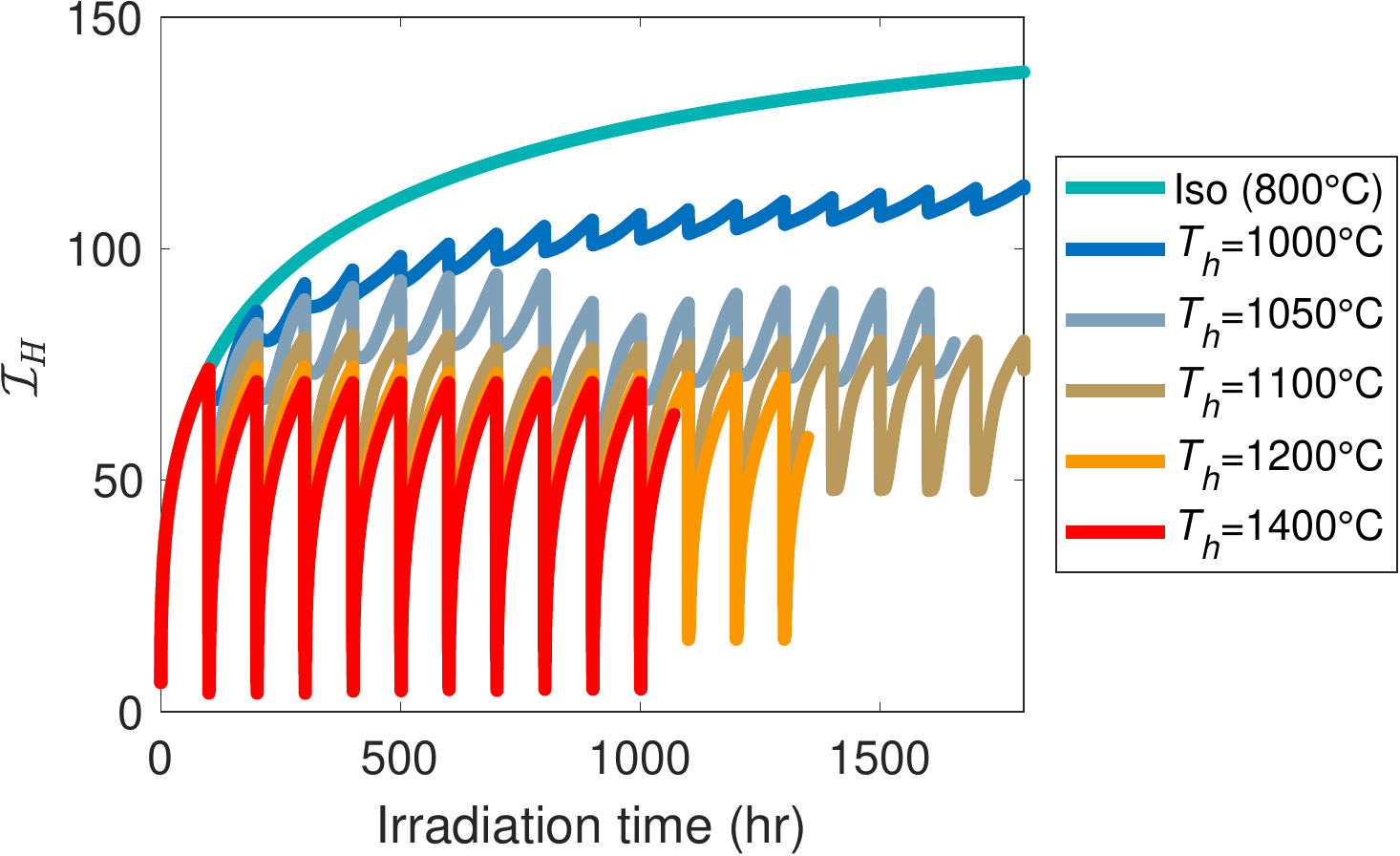}} \hspace{0.05cm}
\subfloat[]{\includegraphics[height=0.20\textheight]{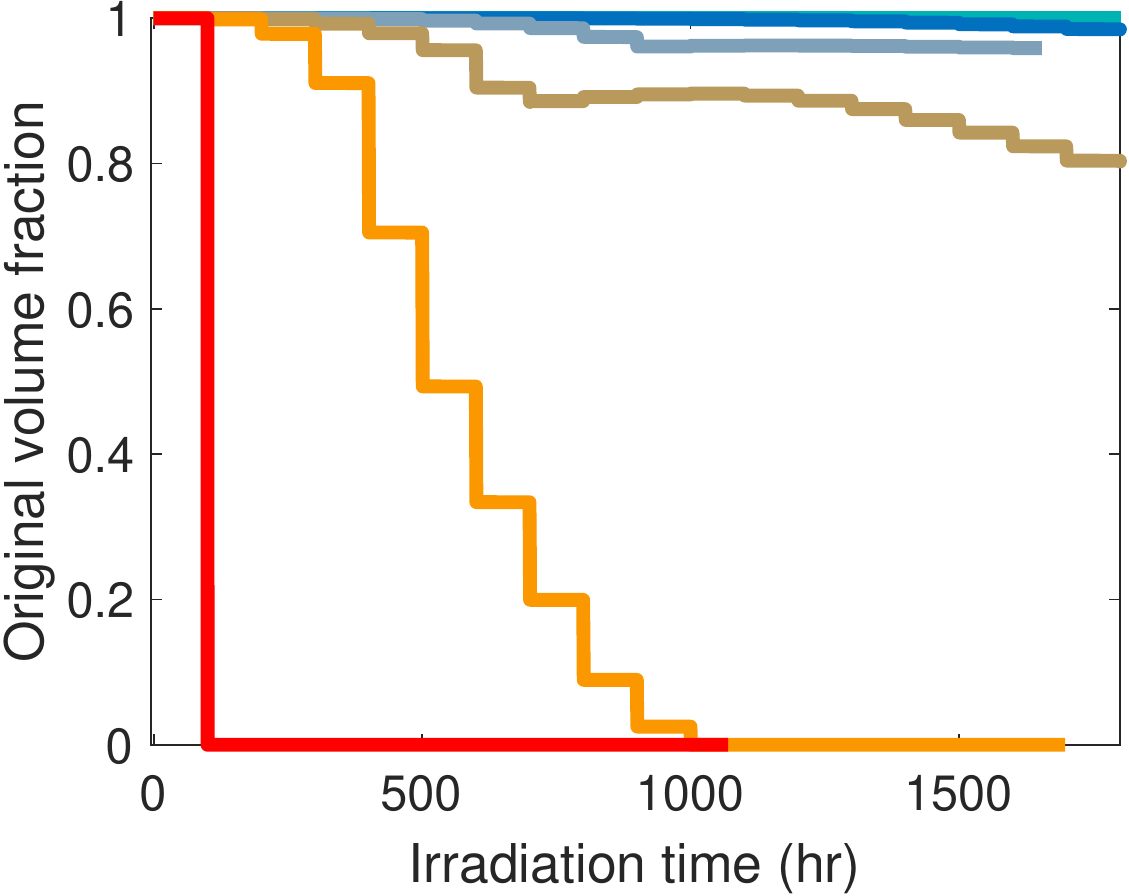}}
\caption{The effect of the annealing temperature $T_h$ on (a) the evolution of the hardness indicator $\mathcal{I}_H$, during annealing every 100 hrs for 1 hr, and (b) the evolution of the original volume fraction.}
\label{fig:varth}
\end{figure}

\paragraph{Temperature of annealing}
In Figure~\ref{fig:varth}a, the influence of the annealing temperature on the hardness indicator is shown, only considering necklace nucleation. An annealing temperature $T_h$ of 1000 \textdegree C during 1 hr after every 100 hrs does not result in full recrystallization, but $\mathcal{I}_H$ is somewhat reduced: during the first 1800 hrs, the maximum value is 113, compared to 138 for the isothermal simulation at 800 \textdegree C. For higher annealing temperatures, of 1050 \textdegree C and 1100 \textdegree C, the maxima are 94 and 82, respectively (for the selected parameter set). For 1200 \textdegree C and 1400 \textdegree C, the maximum is the same, 74. Also, for the annealing temperature of 1000 \textdegree C, the maximum for the hardness indicator keeps increasing, whereas for the higher temperatures, the maximum remains stable.
 Figure~\ref{fig:varth}b displays the evolution of the original volume fraction, which clearly reveals the moment at which no original grains are left in the microstructure. For an annealing temperature of 1400 \textdegree C and higher, full recrystallization takes place during the first annealing cycle, while for 1200 \textdegree C, the original microstructure vanishes only in the 10th cycle. At 1100 \textdegree C or lower, even 18 cycles are not sufficient for full recrystallization to take place. 

\begin{figure}[H]
\subfloat[]{\includegraphics[height=0.22\textheight]{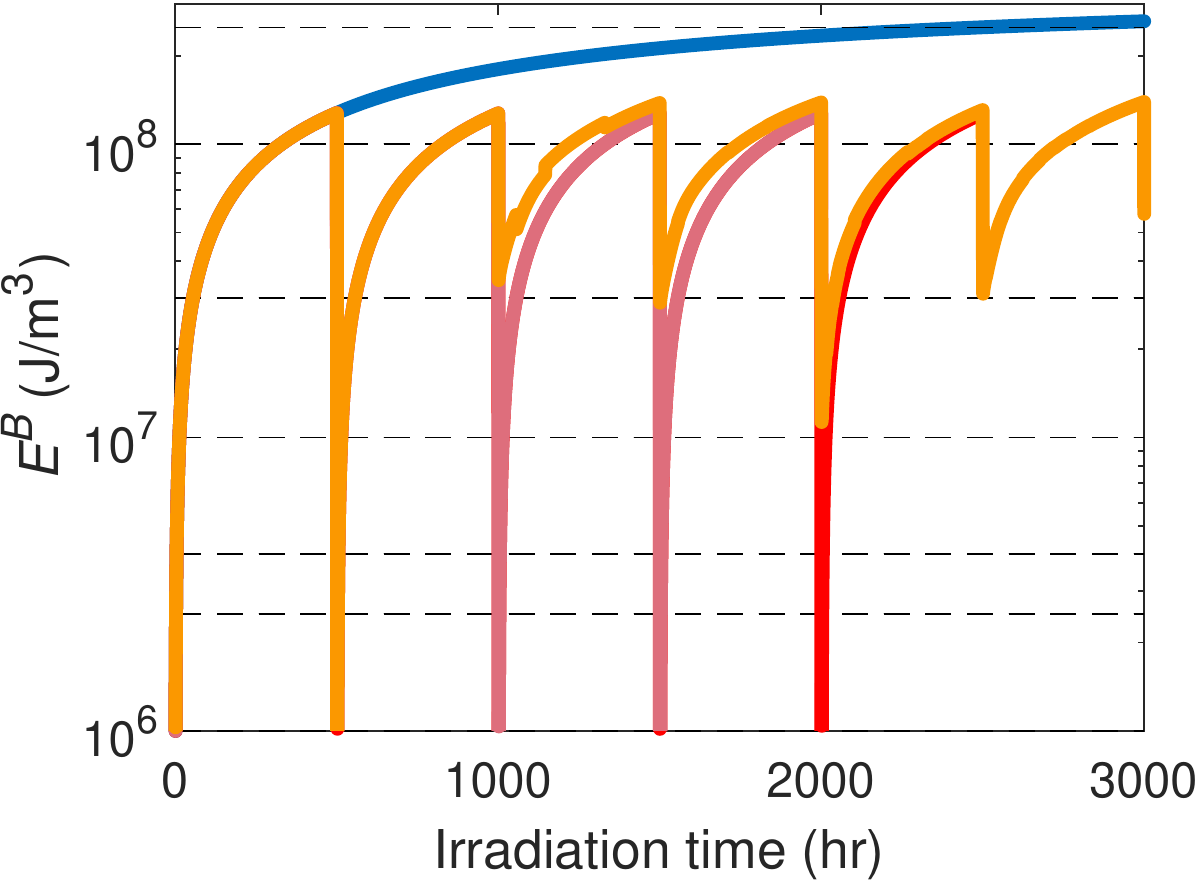}}
\subfloat[]{ \includegraphics[height=0.22\textheight]{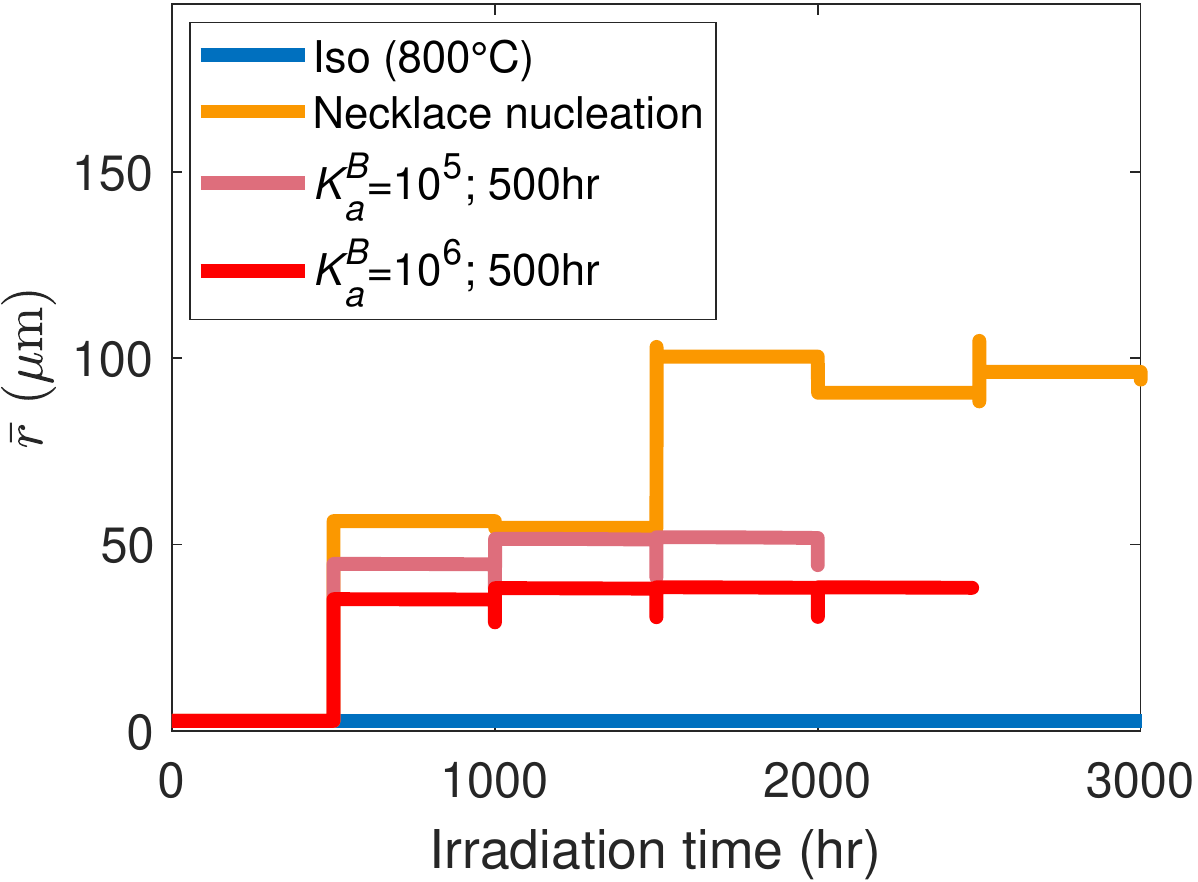}} \vspace{0.01cm}
\subfloat[]{\includegraphics[height=0.22\textheight]{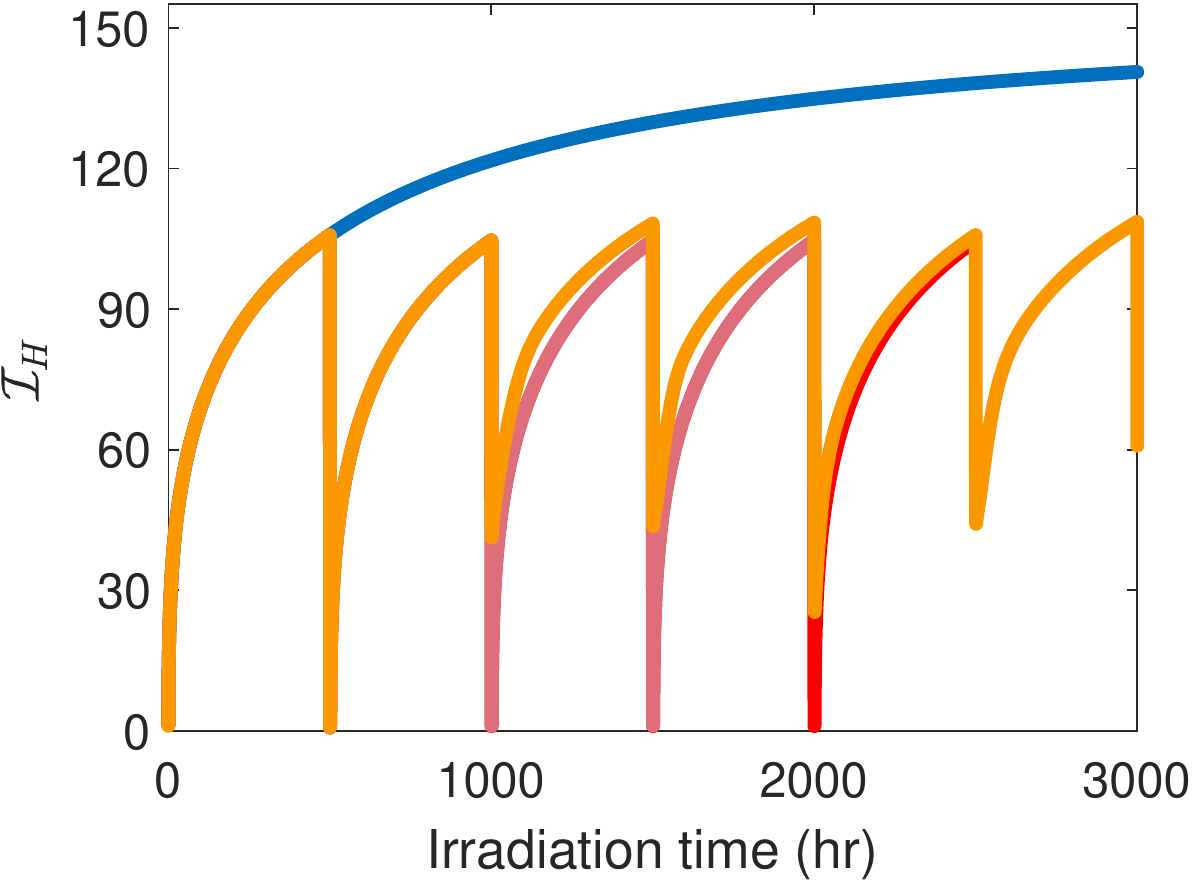}}
\subfloat[]{\includegraphics[height=0.22\textheight]{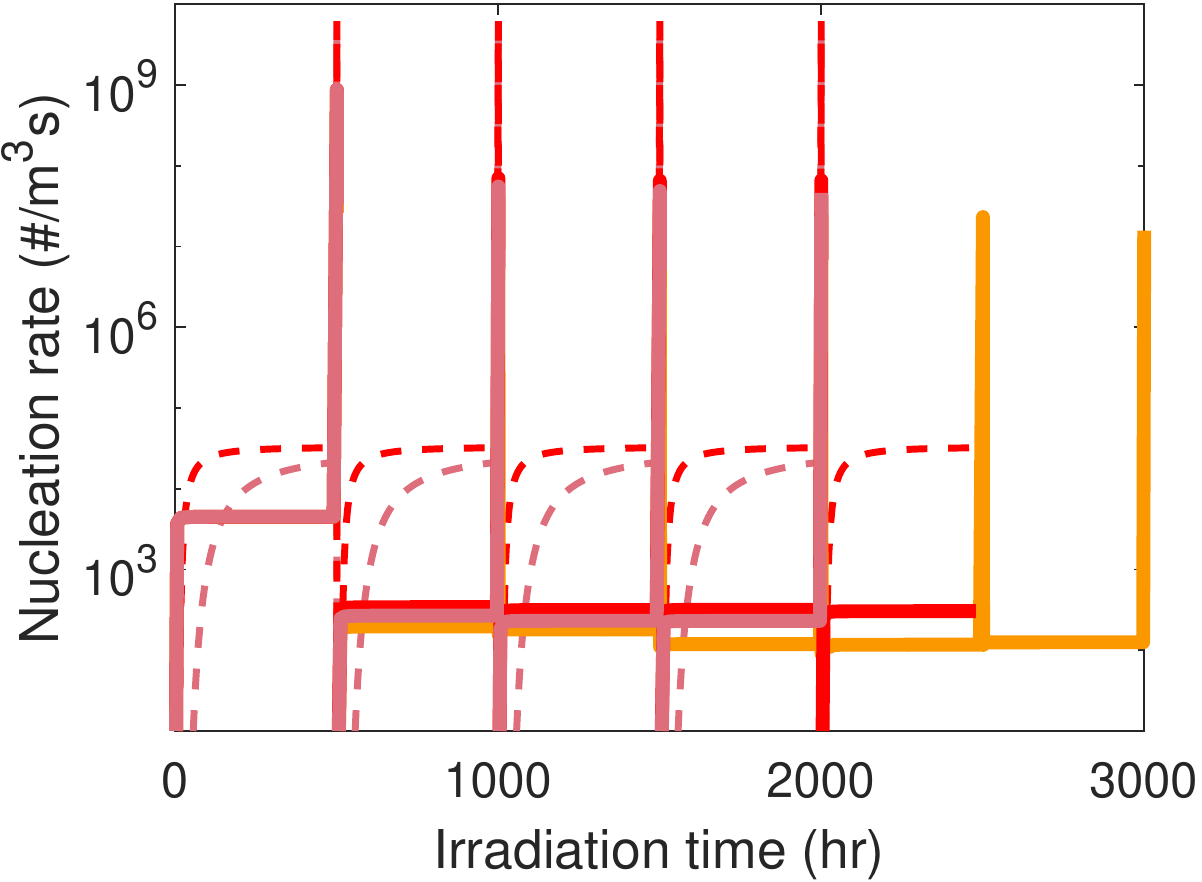}}
\caption{Temporal evolution of (a) the bulk stored energy density $E^B$, (b) the hardness indicator $\mathcal{I}_H$, (c) the average grain radius $\bar{r}$ and (d) the bulk (dashed line) and necklace (solid line) nucleation rate, during irradiation at 800 \textdegree C, with repeated annealing at 1200 \textdegree C every 500 hrs, for different bulk nucleation parameters.}
\label{fig:bulknoniso}
\end{figure}

\paragraph{Bulk nucleation effects}
So far, only necklace nucleation has been considered, as parameterization of the model was done assuming necklace nucleation only. However, the average grain radius rises significantly after the first recrystallization, and bulk nucleation could have a significant effect on the obtained microstructure. Here, the effects for bulk nucleation with $K_N^B=10^{24}$ m$^{-3}$s$^{-1}$ and for two values of $K_a^B$, $K_a^B$=10$^5$ and $K_a^B$=10$^6$, are shown in Figure~\ref{fig:bulknoniso}. 

Bulk nucleation affects the average grain size significantly (Figure~\ref{fig:bulknoniso}b), because of the higher nucleation rate, but the increase of the hardness indicator is only marginally affected by the extra nucleation mechanism (Figure~\ref{fig:bulknoniso}c). The bulk nucleation rate mostly exceeds the necklace nucleation rate. As a general rule, as long as (nearly) full recrystallization is achieved during the annealing period, it does not matter for the hardness indicator how much nucleation takes place. This implies that the more bulk nucleation occurs, the lower the annealing temperature can be, while still achieving full recrystallization.

\section{Conclusions}
The divertor of the nuclear fusion reactors ITER (and DEMO), with plasma-facing parts made out of tungsten, will be heavily damaged during neutron displacement cascades, which leads to strong irradiation hardening, accompanied by a loss of ductility. At sufficiently high temperatures, neutron-induced recrystallization is expected to take place, which (partially) removes the irradiation hardening. At intermediately high temperatures, on the order of 800\textdegree C, recrystallization does not take place easily and interim annealing treatments inducing recrystallization may be an interesting option for regions in the monoblocks where this temperature is not exceeded. \\
\\
In this work, it is explored whether the hardening of neutron-irradiated tungsten might be appreciably reduced by applying repeated cyclic heat treatments. Use is made of a multi-scale model for the microstructural evolution combined with a mean-field recrystallization model (with multiple homogeneous equivalent media) and a cluster dynamics model for the neutron damage. Here, a base temperature of 800 \textdegree C was used, and the annealing heat treatments were applied every 100 hrs or every 500 hrs, during 1 hr at 1200 \textdegree C. The evolution of the hardness was studied qualitatively, using a hardness indicator, which is a function of the concentrations of irradiation defects. For the selected parameter set, it was found that with these treatments, the hardness indicator reduced significantly (here, from more than 140 to less than 70, by heating every 100 hrs, or to 110, by heating every 500 hrs). Increasing the frequency of the cyclic heating can further reduce the value of the hardness indicator. \\
\\
The annealing temperature for the heat treatments was varied from 1000 \textdegree C to 1400 \textdegree C. It was found that the hardness indicator reduces significantly as long as the annealing temperature is chosen sufficiently high to achieve at least partial recrystallization (here, minimum 1050 \textdegree C, for heating every 100 hrs with a base temperature of 800 \textdegree C).\\
\\
Under cascade damage conditions, the stored energy may be high in the grain interior, which possibly entails significant bulk nucleation, whereas for strain-induced recrystallization, nucleation mostly takes place near the grain boundaries. The simulation results show that under these circumstances, a higher total nucleation rate is expected during the heat treatments, leading to a faster or more progressed (partial) recrystallization. As a result, a lower annealing temperature might suffice in the case that bulk nucleation takes place. \\
\\
A systematic parameter characterization of the model was carried out in this study, based on experimental data from static recrystallization experiments. The analysis shows the need for more experimental data, mostly to validate and better capture the temperature-dependency of the microstructural evolution during recrystallization. Furthermore, experimental information is lacking to make an optimal and unique choice for the parameters, which affects the prediction of the grain size evolution. Such experiments should provide temperature-dependent information on the initial and final (average) grain size, initial defect density, as well as on the evolution of the recrystallized fraction. More reliable information on the recrystallization behaviour may also lead to the need to refine the mean-field recrystallization model.\\
\\
The evolution of the hardness indicator that was found here shows that the yield strength of tungsten may vary considerably when the reactor is in operation. Recrystallization might be a way to limit the (maximum) hardness and improve the ductility. Cyclic heat treatments are suggested as a possible route to improve the lifetime of the tungsten monoblocks. In order to apply such heat treatments, a change in the monoblock design might be needed, as the temperature of tungsten close to the cooling tube is now too limited by the maximum temperature of the cooling tube and the cooling fluid.\\
 
\appendix 

\section{Defect evolution details} \label{app:cd}
The full set of cluster dynamics equations (that were summarized by Equation~\ref{eq:gen1} and \ref{eq:gen2}), which are adapted from \cite{Li2012} and \cite{Jourdan2015},  is given below. The largest vacancy cluster size is $N_V$ and the largest interstitial cluster size is $N_I$. 
\subsection{The full Cluster Dynamics equations} \label{app:rn}
\begin{dgroup*}
\begin{dmath*}
\frac{dC_I}{dt} = 	G_I  + k_{I_2+V}^+C_{I_2} C_V + 2 \alpha_2^- C_{I_2}   + \sum_{n=3}^{N_I} \alpha_{n}^- C_{I_{n}} - k_{I+V}^+ (C_I C_V - C_I^{eq} C_V^{eq})  - 2 \alpha_1^+C_I^2   - \sum_{n=2}^{N_I} \alpha_{n}^+ C_I C_{I_{n}}  - \sum_{n=2}^{N_V} k_{V_n+I}^+ C_I C_{V_n} - (k_{D+I}^++k_{S+I}^+) C_I,  
\end{dmath*}
\begin{dmath*}
 \frac{dC_{I_2}}{dt} =   	 G_{I_2} + \alpha_1^+ C_I^2 + k_{{I_3}+ V}^+ C_{I_3} C_V + \alpha_3^- C_{I_3}  + k^-_{I-V} C_I   - \alpha_2^- C_{I_2}  - \alpha_2^+ C_{I_2} C_I - k^-_{I_2-V} C_{I_2} - k^+_{I_2+V}C_{I_2}C_V,
\end{dmath*}
\begin{dmath*}
\frac{dC_{I_{n}}}{dt}=		  G_{I_n}+  \alpha_{n+1}^- C_{I_{n+1}} +  \alpha_{n-1}^+ C_I C_{I_{n-1}}
+ k_{I_{n-1}-V}^- C_{I_{n-1}}  - \alpha_n^{+} C_I C_{I_n} + k_{I_{n+1}+V}^+ C_V C_{I_{n+1}}  - \alpha_{n}^- C_{I_{n}}      -k_{I_{n}+V}^{+} C_V C_{I_n}  - k_{I_n-V}^- C_{I_n}, \qquad \text{ for $3 \le n \le N_I-2,$}
\end{dmath*}
\begin{dmath*}
\frac{dC_{I_{N_{I}-1}}}{dt} =	G_{I_{N_I-1}} +\alpha_{N_I}^- C_{I_{N_I}} +  \alpha_{N_I-2}^+ C_I C_{I_{N_I-2}} + k_{I_{N_I-2}-V}^- C_{I_{N_I-2}}  - \alpha_{N_I-1}^+ C_I C_{I_{N_I-1}}  - \alpha_{N_I-1}^- C_{I_{N_I-1}} -k_{I_{N_I-1}+V}^{+} C_V C_{I_{N_I-1}}   + k_{I_{N_I}+V}^+ C_V C_{I_{N_I}} 	- k_{I_{N_I-1}-V}^- C_{I_{N_I-1}}, 
\end{dmath*}
\begin{dmath*}
\frac{dC_{I_{N_{I}}}}{dt}= 		 G_{I_{N_I}}+ \alpha_{N_I-1}^+ C_I C_{I_{N_I-1}} 	+ k_{I_{N_I-1}-V}^- C_{I_{N_I-1}} - \alpha_{N_I}^- C_{I_{N_I}} -k_{I_{N_I}+V}^{+} C_V C_{I_{N_I}}, 
\end{dmath*}
\begin{dmath*}
 \frac{dC_V}{dt} =   G_V + 2 \gamma_2^- C_{V_2} +k^+_{V_2+I}C_{V_2}C_I  + \sum_{n=3}^{N_V} \gamma_{n}^- C_{V_{n}}  + \sum_{n=2}^{N_{I+1}} k_{I_{n-1}-V}^- C_{I_{n-1}} - k_{I+V}^+ (C_I C_V - C_I^{eq} C_V^{eq})  - 2 \gamma_1^+ C_V^2 - \sum_{n=2}^{N_V} \gamma_n^+ C_V C_{V_n}  - \sum_{n=2}^{N_I}  k_{I_n+V}^+ C_V C_{I_n}  -    (k_{D+V}^++k_{S+V}^+)C_V, 
\end{dmath*}
\begin{dmath*}
 \frac{dC_{V_{n}}}{dt} = 	G_{V_n} +   k_{V_{n+1}+I}^+ C_I C_{V_{n+1}} + \gamma_{n+1}^-  C_{V_{n+1}}   - k_{V_n+I}^+ C_I C_{V_n} + \gamma_{n-1}^+ C_V  C_{V_{n-1}}
 							   - \gamma_n^- C_{V_n} - \gamma_n^+ C_V C_{V_{n}}, \qquad \hspace{7cm} \text{ for $2 \le n \le N_V-2,$}
\end{dmath*}
\begin{dmath*}
 \frac{dC_{V_{N_V-1}}}{dt} =  		  G_{N_{N_V-1}} + k_{V_{N_V}+I}^+ C_I C_{V_{N_V}} + \gamma_{N_V}^-  C_{V_{N_V}} 
								+ \gamma_{N_V-2}^+ C_V  C_{V_{N_V-2}}
 							  - k_{V_{N_V-1}+1}^+ C_I C_{V_{N_V-1}}  - \gamma_{N_V-1}^- C_{V_{N_V-1}}
  							- \gamma_{N_V-1}^+ C_V C_{V_{N_V-1}},
\end{dmath*}
\begin{dmath*}
 \frac{dC_{V_{N_V}}}{dt} = 			  G_{N_{N_V}} + \gamma_{N_V-1}^+ C_V  C_{V_{N_V-1}} 
 							  - k_{V_{N_V}+I}^+ C_I C_{V_{N_V}}  - \gamma_{N_V}^- C_{V_{N_V}} .
\end{dmath*}
\end{dgroup*}
The formulas for computation of each rate coefficient can be found in Table \ref{tab:ratecoef}, and the calculation of several parameters in this table is specified hereafter. The parameter values that have been used can be found in Table \ref{tab:parval}. The expressions related to the evolution of the dislocation density are detailed in section~\ref{app:dis}.

\begin{table*}[ht!]
	\caption{Rate coefficients. The superscript `+ ` denotes absorption of a point defect and the subscript `-' denotes emission.} 
	\label{tab:ratecoef}
\centering
	\begin{tabular}[ht!]{ >{$}r<{$} |  >{$}r<{$} | l }
	 \multicolumn{2}{c}{\bf{Dislocation loop \(\mathbf{ I_n} \) }} \\ 
	\textit{Absorption rate}					&	\textit{Emission rate} 													&\\ \hline
	\alpha_n^+ = 2 \pi r_{I_n}Z_{I_n}^I D_I 		&  \alpha_n^- =  2 \pi r_{I_{n-1}}Z_{I_{n-1}}^I D_I \text{ exp } ( - E_{I_n-I}^b / k_B T ) / V_{at}	 		& Interstitial\\
	k_{I_n+V}^+ = 2 \pi r_{I_n}Z_{I_n}^V D_V		& k_{I_{n-1}-V}^- =  2 \pi r_{I_{n-1}}Z_{I_{n-1}}^V D_V \text{ exp } ( - E_{I_n-V}^b / k_B T ) / V_{at}		& Vacancy \\
	k^+_{I+V} = 4 \pi r_{IV} (D_I + D_V)   		& 			-														& VI-recombination \\
	\multicolumn{2}{c}{\bf{Vacancy cluster  \(\mathbf{V_n} \)}} \\
	\textit{Absorption rate}					&	\textit{Emission rate} 													&\\ \hline
	k_{V_n+I}^+ = 4 \pi r_{V_n} D_I 			& 	-																& Interstitial\\
	\gamma_n^+ = 4 \pi r_{V_n} D_V			& \gamma_n^- = 4 \pi r_{V_{n-1}} D_V \text{ exp } \big( -E^b_{V_n-V} / k_B T\big)	/ V_{at}			& Vacancy \\
	\end{tabular}
\end{table*}

\begin{table*}[ht!]	
	\caption{Rate coefficients related to the strengths of the sinks (grain boundaries and dislocations).} 
	\label{tab:rc2}
	\begin{tabular}[ht!]{ >{$}r<{$}|  l}
\multicolumn{2}{c}{\textbf{Dislocation sink \( \rho_D \)}} \\ \hline
	k_{D+I}^+ = \rho Z_D^I D_I 				& Interstitial \\									
	k_{D+V}^+ = \rho Z_D^{V} D_{V} 			& Vacancy \\
	& \\
	\multicolumn{2}{c}{\textbf{Grain boundary sink}}  \\  \hline
 	k_{S+I}^+ = 3 S_I^{sk} D_I / r_{grain} & Interstitial\\
 	k_{S+V}^+ = 3 S_V^{sk} D_V / r_{grain} & Vacancy \\
\end{tabular}
\end{table*}

The sink strength sums of Table~\ref{tab:rc2} are given by: 
\begin{align}
\bigg( S_{I}^{sk} \bigg)^2	&	=  \frac{1}{D_I} \bigg[ \sum_{n=1}^{N_I-1} \alpha^+_n C_{I_n} + \sum_{n=1}^{N_V}  k^+_{V_n+I} C_{V_n} \bigg] + \rho Z_D^I, \\
\bigg(S_{V}^{sk} \bigg)^2 &		=    \frac{1}{D_V} \bigg[ \sum_{n=1}^{N_V-1}  \gamma^+_n C_{V_n} + \sum_{n=1}^{N_I} k^+_{I_n+V} C_{I_n} \bigg] + \rho Z_D^V.
\end{align}The binding energies, that are used in the expressions for the emission rates are calculated using the capillarity approximation, using \cite{Li2012}: 
\begin{align}
E^f_{I_n} & = E^f_{I_{n-1}}+E^f_{I} -E^b_{I_{n}-I} \text{ (by definition)}, \\
E_{I_n-I}^b & =E_I^f + \frac{E_{I_2}^b-E_I^f}{2^{2/3}-1} \big[ n^{2/3} - (n-1)^{2/3} \big],  \\
E_{I_n-V}^b & =E_V^f + \frac{E_{I}^f-E_{I_2}^b}{2^{2/3}-1} \big[ n^{2/3} - (n-1)^{2/3} \big], \\
E_{V_n-V}^b &=E_V^f + \frac{E_{V_2}^b-E_V^f}{2^{2/3}-1} \big[ n^{2/3} - (n-1)^{2/3} \big]. 
\end{align}The diffusion coefficients of the isolated point defects $I$ and $V$ are given by: 
\begin{align}
D_I= & D_{I_0} \mathrm{exp}(-E_I^m/k_B T), \\
D_V= & D_{V_0} \mathrm{exp}(-E_V^m/k_B T).
\end{align}The dislocation bias factors, used in the dislocation network sink rate and the absorption rates for vacancies and self-interstitial atoms by self-interstitial defect clusters are: 
\begin{align}
Z_{I_n}^{I} = & Z_D^{I} \mathrm{max} \big[ \frac{2 \pi}{\ln(8 r_{I_n}/r_p)},1 \big], \\	
Z_{I_n}^{V} = & Z_D^{V} \mathrm{max} \big[ \frac{2 \pi}{\ln(8 r_{I_n}/r_p) },1\big], \\		
\end{align}where $r_p = 2b$ is assumed, following \cite{Li2012}. The capture radii $r_{I_n}$ and $r_{V_n}$, for interstitial and vacancy clusters of size $n$ are given by: 
\begin{align}
r_{I_n} = & \sqrt{\frac{nV_{at}}{\pi b}}, \label{eq:In}\\ 
r_{V_n} = & (3nV_{at}/4\pi)^{1/3} + \sqrt{3}a_0/4. \label{eq:Vn}
\end{align}

\subsection{Dislocation evolution details} \label{app:dis}
In the evolution of the dislocation density (Equation \ref{eq:gen3}) \cite{Stoller1990}, the climb velocity of the dislocations is given by:
\begin{align}
v_{cl}=\frac{2 \pi}{b \ln{R/r_c}} \bigg[ Z^I_D D_I C_I - Z^V_D D_V (C_V - C_V^D) \bigg],
\end{align}where, $C_V^D$ is the equilibrium concentration of vacancies near a dislocation, $C_V^D = C_V^{eq} \exp{\bigg( \frac{\sigma V_{at}}{k_B T}\bigg)}$ \cite{Was2007}, where $C_V^{eq}$ is the equilibrium vacancy concentration in the bulk, $V_{at}$ is the atomic volume, $\sigma$ is the internal stress due to pinned dislocations, $\sigma = A \mu b \sqrt{\rho_p}$, where $\rho_p = 0.1 \rho$ is the pinned dislocation density and $A$=0.4. $R/r_c=2 \pi$ is taken, where $R$ and $r_c$ are the outer radius and core radius of the dislocations. \\
\\
Furthermore, $S_{BH}$, the Bardeen-Herring source density is given by $S_{BH}=\big( \rho_p/3 \big)^{1.5}$ and $d_{cl} = \big( \pi \rho \big)^{-1/2}$ is the distance that dislocations can travel before they annihilate \cite{Stoller1990}. 

\subsection{Model parameters} \label{app:par}
\begin{table}[H]
\centering
\caption{Parameter values.}
\label{tab:parval}
	\begin{tabular}{>{$}l<{$} | l      |       l                |            l              |  r}
	\textbf{Parameter}		&	\textbf{Unit}	& \textbf{Value} & \textbf{Description}		&	\textbf{Reference}						\\ \hline
	a_0			&	nm	&	0.31652	& Lattice parameter 			& \cite{Lassner2012} 				\\
	\alpha_T	& - 	& 	0.15 	& Taylor barrier strength 		& \cite{Davoudi2014} \\
	\gamma_b 		&	J/m$^2$	&	0.869		& GB-surface energy		& \cite{Lopez2015} 	\\
	\delta			& 	nm 	&	1		& grain boundary thickness			& \cite{Favre2013}   	\\	
	D^{GB}_0		&	m$^2$/s	&	0.27\e{-4}	& Self-diffusivity 	& Estimated, \\	
				& 			&  			& along grain boundaries&  using \cite{Lassner1999}		\\		
	D_{I_0}		&	m$^2$/s	&	8.77\e{-8}	& SIA-diffusivity	& \cite{Faney2013}	          \\
	D_{V_0}		&	m$^2$/s	&	177\e{-8}	& Vacancy diffusivity	& \cite{Faney2013}		\\	
	E^f_I 			&	eV	&	9.466		& Formation energy SIA		& \cite{Li2012}		\\
	E^f_V			&	eV	&	3.80		& Formation energy vacancy	& \cite{Li2012}		\\
	E^b_{I_2}		&	eV	&	2.12		& Binding energy SIA-SIA		& \cite{Li2012}		\\
	E^b_{V_2}		&	eV	&	0.6559	& Binding energy V-V		& \cite{Li2012}		\\
	E^m_I 		&	eV	&	0.013		& SIA migration energy 		& \cite{Faney2013}		\\
	E^m_V 		&	eV	&	1.66		& Vacancy migration energy 	& \cite{Faney2013}		\\
	K_a^A   		& 	- 	&	5\e{7}		& Nucleation activation energy reduction & -			\\
	K_m 			&	- 	& 	1490		& grain boundary mobility parameter 		& -				\\
	K_N 			&	\#/m$^2$s &	2.5\e{17}	& Nucleation rate constant	& -				\\	
	M 			& - 		&  	3.06   		& Taylor factor 			& \cite{Stoller2000} 	\\
	\mu			&	Pa 	&	161\e{9} 	& Shear modulus	&  \cite{Lassner1999}	\\
	Q^{GB}		&	J/mol	&	4\e{5}	& Activation energy for GB mobility	& \cite{Lassner1999}				\\
	r_{IV}			&\r{A}   	&       4.65               & Recombination radius		& \cite{Li2012}                 \\       
	V_m 			&  m$^3$/mol & 9.55\e{-6} 	& Molar volume 	& - 				\\ 
   	Z_D^I 		&	- 	& 	1.2		& SIA-dislocation bias		&  \cite{Li2012}		\\
	Z_D^V		& 	- 	&	1		& V-dislocation bias			&	\cite{Li2012}		\\	
	\end{tabular}
\end{table}

\paragraph{Simulation settings}\mbox{}\\
The simulations were performed using 17 different HEMs, with the following 16 limits for the bulk stored energy density that cover a range from $1\e2$ J/m$^3$ to $2.5\e8$ J/m$^3$ (namely:  1\e{2};  1\e{3}; 3\e{3}; 6\e{3}; 1\e{4}; 5\e{4}; 1\e{5}; 2.5\e{5}; 5\e{5}; 1\e{6}; 2.5\e{6}; 4\e{6}; 1\e{7}; 3\e{7}; 1\e{8}; 2.5\e{8}). The number of representative grains at the start of a simulation was taken to be 16. Besides those representative grains, each HEM was allowed to have up to 20 representative grains that were nucleated in the course of the simulation, so in total the maximum number of representative grains amounted up to 370, sufficient to allow for a distribution of grain sizes within each HEM. In some simulations, the maximum number of representative grains per HEM was taken to be 8 (for bulk nucleation, with $K^B_a=10^5$ and $K^B_a=10^6$, and for necklace nucleation with extra heating every 100 hrs) or 12 (for necklace nucleation with annealing every 500 hrs). In all simulations, the nucleation threshold was taken to be $E^{B,thr}_{nuc}=10^6$ J/m$^3$. In the cluster dynamics model, $N_I$ = $N_V$ = 100 is used for all simulations. The simulations that were performed in section~\ref{sec:parkar} were performed with 20 different HEMs, using 3 extra HEM limits (namely: 1.3\e{6}, 1.6\e{6} and 2.0\e{6}).

\section{Solution procedure}\label{sec:sp}
The solution procedure is based on \cite{Mannheim2018b}, but with several modifications to account for non-isothermal loading and for multiple HEMs.
\begin{enumerate}
\item \textbf{Time step.} The size of the time step $\Delta t = t_{i+1} - t_i$ is calculated, where the constraints of Table~\ref{tab:ts} are taken into account. 
\item \textbf{Defect evolution.} The evolution of the defect concentrations for each grain is calculated using cluster dynamics. 
\item \textbf{Redistribution.} Based on $E^B$, the representative grains are assigned to the HEMs. If the maximum amount of representative grains in a HEM is exceeded, then the two representative grains in this HEM that are most alike (based on their bulk and surface properties $E^B$ and $E^S$) are merged to form one representative grain.
\item \textbf{Nucleation.} For necklace nucleation, $\dot{N}^A$, $r^{A}_{nuc}$, $E^A_{act}$ and $A_{nuc}$ are calculated. The volume for the newly nucleated grains is delivered by all the HEMs, according to the surface fractions. Within a HEM, the nucleated volume is provided by the representative grains according to their volume fraction. In the simulations where bulk nucleation is taken into account, $\dot{N}^B$, $r^{B}_{nuc}$, $E^{B}_{act}$ and $V_{nuc}$ are calculated. Also for bulk nucleation, all representative grains that contribute to the nucleation volume, deliver the volume for the new nuclei according to their volume fractions. The nuclei that are formed are placed in the lowest energy HEM. A HD-daughter grain can only merge with other HD-daughter grains.
\item \textbf{Time step check.} If more grains nucleated in the previous time increment than were already present (counting nucleated grains only), then the calculations for this time increment are repeated, using a reduced time step size $\Delta t$. 
\item \textbf{Grain growth.} All regular grains obtain new grain radii and new concentrations. If grains vanish, then subincrements are used to avoid negative radii. 
\item \textbf{Time step check.} If the overall stored energy density of the microstructure decreased too much in a single time increment (more than 5\%), then the calculations are repeated with a smaller time step size $\Delta t$.
\end{enumerate}

\paragraph{Volume conservation during grain growth}\mbox{}\\
In general, the volume change of a grain $k$ during grain growth is calculated using Equation~\ref{eq:dv}. An exception is made for grains that shrink with respect to their own HEM. If a grain $k$ in HEM $q$ shrinks with respect to HEM $q$, then the appropriate term in the summation of Equation~\ref{eq:dv} is replaced by the following:
\begin{align}
\Delta V^{q}_k = - 4 \pi m \Delta t r_k^2 \phi^{q} \Delta E^{q}_k \frac{\sum_{k \in q: \Delta V^{q}_k >0} \Delta V_k^{q} N_k}{\sum_{k \in q: \Delta V^{q}_k <0} \Delta V_k^{q} N_k},
\end{align}to conserve the volume, after \cite{Bernard2011}.

\begin{table}[H]
	\begin{tabular}{|p{2 cm}|p{11 cm}|}
	\hline
	Mechanism & Rule \\ \hline
	General & The size of the time step can increase no more than 5\% with respect to the step size during the previous time increment (or, when the step size is smaller than 10 s, an increase of up to 50\% is allowed). \\ \hline
	Grain growth & No grain is allowed to growth or shrink more than 10\% with respect to a HEM (unless the grain is smaller than 0.1\% of the average grain volume). This is estimated based on the relative volume changes during the previous time increment. \\ \hline
	Nucleation & The amount of nucleated grains $N_{new}$ during a time step may not exceed the total amount of newly formed grains $N_{new,tot}$ that already exist in the microstructure. If this number is exceeded, then all events during the time increment are recalculated, using an adjusted time step size $dt \rightarrow dt \times N_{new,tot}/N_{new}$. \\ \hline
	Temperature & During a single time increment, the temperature may increase no more than 30 K. Also, the starts and ends of the dwell times should coincide with the start/end of a time increment. Since the temperature is prescribed, these adjustments can be made at the beginning of the time increment.\\ \hline
	Recovery & The total energy density $E=\sum_{n=1}^{nHEMs} \gamma^{n} E^{HEM,n}$ should not decrease more than 5\% during a single time increment: \newline if $(E_{i+1} - E_i)/E_i > -0.05$, \newline then $dt \rightarrow dt / (|E_{i+1} - E_i| / 0.05E_i)$, \newline where $E(i)$ is the total energy density during the previous time increment; and $E(i+1)$ the total energy density during the first try for the current time increment. \\ \hline
	\end{tabular}
	\caption{Time step limitations.}
	\label{tab:ts}	
\end{table}


\paragraph{Gibbs free energy change during nucleation}\mbox{}\\
To determine the nucleation radii and activation energies for necklace nucleation and bulk nucleation, the following expressions are used:
\begin{align}
\frac{\partial{\Delta E}}{\partial{t}} & = \frac{\partial \Delta E }{\partial E^{B}}\frac{d E^{B}}{dt} + \frac{\partial \Delta E}{\partial r}\frac{dr}{dt} = \\
&  = \frac{1}{K^A_a} \bigg[ - \frac{4 \pi}{3}r^3 \frac{d E^B}{dt } - 4 \pi r^2 m (E^B-E^B_0) (E - E^B_0 ) + 6 \pi r m \gamma_b  (E^B + E - 2E^B_0 ) - 9 \pi m \gamma_b^2  \bigg] < 0,
\intertext{for necklace nucleation, and}
\frac{\partial{\Delta E}}{\partial{t}} & = \frac{\partial \Delta E }{\partial E^{B,HD}}\frac{d E^{B,HD}}{dt} + \frac{\partial \Delta E}{\partial r}\frac{dr}{dt}  =\nonumber \\
& = \frac{1}{K^B_a} \bigg[ - \frac{4 \pi}{3}r^3 \frac{d E^{B,HD}}{dt } - 4 \pi r^2 m (E^{B,HD}-E^B_0) (E^{HD} - E^B_0 )  \nonumber \\
	& + 6 \pi r m \gamma_b  (E^{B,HD} + \frac{4}{3} E^{HD} - \frac{7}{3} E^B_0 ) - 12 \pi m \gamma_b^2  \bigg] < 0,
\end{align}for bulk nucleation. Here, $E^{B,HD}$ and $E^{HD}$ are the (volume) average stored energy densities of the representative grains for which $E^B$ exceeds the nucleation threshold energy density. As detailed in \cite{Mannheim2018b}, the derivatives are determined numerically. The nucleated grain size is taken to be $r_{nuc}=1.01r_0$, where $r_0$ denotes the largest solution to the above equations. The nucleation activation energy follows from $E_{act} = \Delta E(r^*)$, where $r^*$ is the solution to the static case, $d \Delta E/dr = 0$. 

\section{SRX parameter characterization}\label{app:srx}
\begin{figure}[ht!]
\includegraphics[width=0.6\textwidth]{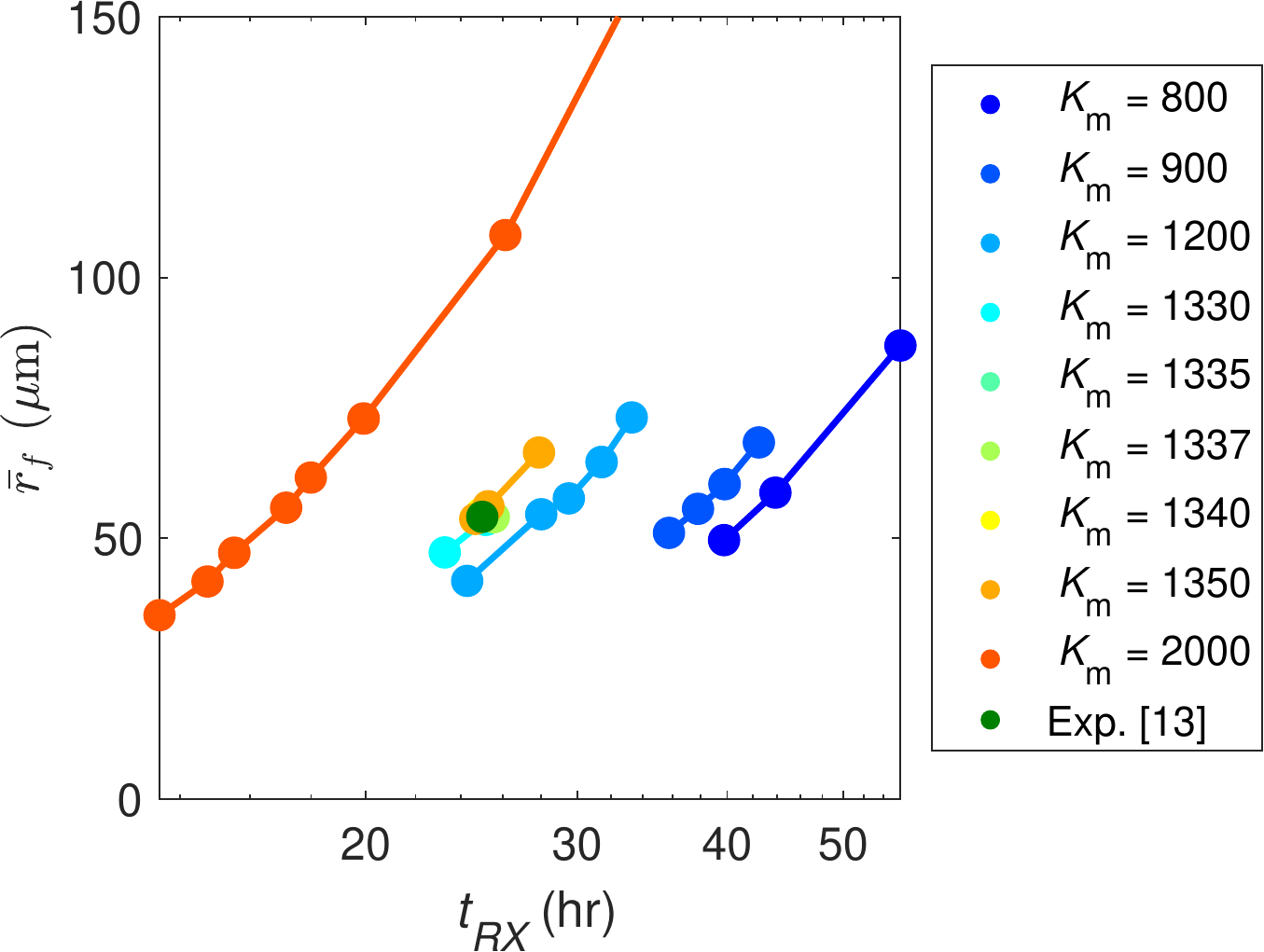}
\caption{Final grain size $\bar{r}_f$ after static recrystallization and time of full recrystallization $t_{RX}$ for $K_a$=3\e{6} and for various values of $K_m$ and $K_N$, in comparison to the experimental result of \cite{Lopez2015}.}
\label{fig:srx}
\end{figure}

Figure~\ref{fig:srx} shows the dependence of the final average grain size and the time to complete recrystallization on the parameters $K_N$ and $K_m$ for a given $K_a$, for static recrystallization, using the microstructure parameters, process conditions and simulation settings as mentioned in section~\ref{sec:parkar}. 

\end{document}